\documentclass[12pt]{article}
\usepackage{amsmath}
\usepackage{mathrsfs}
\usepackage{url}

\usepackage{epsfig,morefloats}
\usepackage{amsfonts}
\usepackage{natbib}
\usepackage{multirow}
\usepackage{enumitem}
\usepackage{amsthm}
\usepackage{amssymb}
\usepackage{color}
\usepackage{lscape}
\usepackage{caption}
\usepackage{subfigure}
\usepackage{graphicx}
\usepackage{float} 
\usepackage{mathabx}
\usepackage{comment} 
\usepackage{multirow}

\usepackage{pdflscape}  % 提供横向页面支持
\usepackage{booktabs}   % 专业表格线
\usepackage{tabularx}   % 自动调整列宽
\usepackage{array}      % 增强表格功能
\usepackage{diagbox}

\usepackage[figuresright]{rotating}

\newcommand{\Date}[1]{\def\@Date{#1}}
\def\today{\number\day~\ifcase\month\or
	January\or February\or March\or April\or May\or June\or
	July\or August\or September\or October\or November\or December\fi~\number\year}

\def\be{\begin{equation}}
	\def\ee{\end{equation}}
\def\bea{\begin{eqnarray}}
	\def\eea{\end{eqnarray}}
\def\bd{\begin{displaymath}}
	\def\ed{\end{displaymath}}
\def\bda{\begin{eqnarray*}}
	\def\eda{\end{eqnarray*}}
\def\bsm{\begin{small}}
	\def\esm{\end{small}}

\def\t0{\theta_0}
\def\nn{\nonumber}

\def\ha1{\widehat \beta_1}

\newcommand{\bone}{\mbox{\bf 1}}

\def\bnt{\begin{enumerate}}
	\def\ent{\end{enumerate}}
\def\T{{ \mathrm{\scriptscriptstyle T} }}
\def\A{{ \mathrm{\scriptscriptstyle A} }}

\def\W{{ \mathrm{\scriptscriptstyle W} }}
\def\AW{{ \mathrm{\scriptscriptstyle AIPW} }}
\def\EL{{ \mathrm{\scriptscriptstyle HC} }}
\def\AEL{{ \mathrm{\scriptscriptstyle HDC} }}
\def\MDEL{{ \mathrm{\scriptscriptstyle MHDC} }}

\def\HC{{ \mathrm{\scriptscriptstyle HC} }}
\def\HDC{{ \mathrm{\scriptscriptstyle HDC} }}
\def\MHDC{{ \mathrm{\scriptscriptstyle MHDC} }}

\def\opt{{ \mathrm{\scriptstyle opt} }}
\def\oc{{ \mathrm{\scriptstyle oc} }}
\def\ps{{ \mathrm{\scriptstyle ps} }}
\def\ini{{ \mathrm{\scriptstyle ini} }}

\def\bsc{\begin{scriptsize}}
	\def\esc{\end{scriptsize}}

\newtheorem{theorem}{Theorem}

\newtheorem{proposition}{Proposition}

\theoremstyle{definition}
\newtheorem{as}{Condition}
\newtheorem{ex}{Example}

\newcommand{\E}{\mathbb{E}}
\newcommand{\PP}{\mathbb{P}}

\makeatletter
\newcommand{\figcaption}{\def\@captype{figure}\caption}
\newcommand{\tabcaption}{\def\@captype{table}\caption}
\makeatother

\newcommand{\AR}{{\rm AR}}

\newcommand{\argmin}{{\rm argmin}}

\newcommand{\diag}{{\rm diag}}

\newcommand{\tr}{\mbox{tr}}
\newcommand{\var}{\mbox{Var}}

\newcommand{\bA}{{\mathbf A}}

\newcommand{\bI}{{\mathbf I}}
\newcommand{\bJ}{{\mathbf J}}

\newcommand{\bM}{{\mathbf M}}

\newcommand{\bS}{{\mathbf S}}
\newcommand{\bU}{{\mathbf U}}
\newcommand{\bV}{{\mathbf V}}
\newcommand{\bW}{{\mathbf W}}
\newcommand{\bX}{{\mathbf X}}
\newcommand{\bY}{{\mathbf Y}}
\newcommand{\bZ}{{\mathbf Z}}
\newcommand{\ba}{{\mathbf a}}
\newcommand{\bb}{{\mathbf b}}
\newcommand{\bfd}{{\mathbf d}}

\newcommand{\bfe}{{\mathbf e}}
\newcommand{\bff}{{\mathbf f}}

\newcommand{\bw}{{\mathbf w}}
\newcommand{\bx}{{\mathbf x}}

\newcommand{\bbeta}  {\boldsymbol{\beta}}

\newcommand{\blambda}{\boldsymbol{\lambda}}

\newcommand{\bSigma}{\boldsymbol{\Sigma}}

\newcommand{\bgamma}{\boldsymbol{\gamma}}

\newcommand{\bpsi} {\boldsymbol{\psi}}

\newcommand{\btheta} {\boldsymbol{\theta}}
\newcommand{\bxi} {\boldsymbol{\xi}}
\newcommand{\bmu} {\boldsymbol{\mu}}

\newcommand{\bC}{{\mathbf C}}

\newcommand{\bzero}{{\mathbf 0}}

\usepackage{accents}

\newcommand{\blind}{1}

\usepackage{xr}
\makeatletter
\newcommand*{\addFileDependency}[1]{
  \typeout{(#1)}
  \@addtofilelist{#1}
  \IfFileExists{#1}{}{\typeout{No file #1.}}
}
\makeatother
\newcommand*{\myexternaldocument}[1]{%
    \externaldocument{#1}%
    \addFileDependency{#1.tex}%
    \addFileDependency{#1.aux}%
}
\myexternaldocument{SM}

\begin{document}
	
	\def\spacingset#1{\renewcommand{\baselinestretch}%
	{#1}\small\normalsize} 
    \spacingset{1} 

    \if1\blind
    {
    \title{\bf Multi-source Learning for Target Population by High-dimensional Calibration}
    \author{Haoxiang Zhan
    \hspace{.2cm}\\
     School of Mathematical Sciences, Peking University, Beijing, China\\
    and \\
    Jae Kwang Kim \\
    Department of Statistics, Iowa State University, Iowa, USA \\ 
    and \\
    Yumou Qiu \\
    School of Mathematical Sciences, Peking University, Beijing, China}
    \date{}
    \maketitle
    } \fi
        
    \if0\blind
    {
    \bigskip
    \bigskip
    \bigskip
    \begin{center}
    {\LARGE\bf Multi-source Targeted Learning by High-dimensional Calibration}
    \end{center}
    \medskip
    } \fi
        
    \bigskip
    \begin{abstract}
    Multi-source learning is an emerging area of research in statistics, where information from multiple datasets with heterogeneous distributions is combined to estimate the parameter of interest for a target population without observed responses. 
    We propose a high-dimensional debiased calibration (HDC) method and a multi-source HDC (MHDC) estimator for general estimating equations. The HDC method uses a novel approach to achieve Neyman orthogonality for the target parameter via high-dimensional covariate balancing on an augmented set of covariates. It avoids the augmented inverse probability weighting formulation and leads to an easier optimization algorithm for the target parameter in estimating equations and M-estimation.
    The proposed MHDC estimator integrates multi-source data while supporting flexible specifications for both density ratio and outcome regression models, achieving multiple robustness against model misspecification. Its asymptotic normality is established, and a specification test is proposed to examine the transferability condition for the multi-source data. 
    Compared to the linear combination of single-source HDC estimators,
    % the linear combination of the DEL estimators obtained using each source separately, 
    the MHDC estimator improves efficiency by jointly utilizing all data sources. 
    % by leveraging multi-source data in a unified procedure.  
    Through simulation studies, we show that the MHDC estimator accommodates multiple sources and multiple working models effectively and performs better than the existing doubly robust estimators for multi-source learning. An empirical analysis of a meteorological dataset demonstrates the utility of the proposed method in practice. 
    \end{abstract}
        
    \noindent%
    \vfill

    \newpage
    % \spacingset{1.9} % DON'T change the spacing!

	\section{Introduction}

    Multi-source learning has become increasingly important in scientific research when multiple datasets are observed at different times, locations, or conditions. These datasets often exhibit distributional heterogeneity but share underlying structural similarities, allowing researchers to extract more information than any single dataset could provide \citep{Li2021TransLasso, Li2023Datafusion, Guo2024Maximin}.
    %addressing the fundamental challenge on how to exploit multiple data sources with heterogeneity for learning tasks in the target domain. 
    %Multi-source learning not only enhances the robustness and generalizability of findings but also maximizes the utility of existing data resources.
    % Integrating multiple sources of data gives more information to learn the parameter of interest on the target population. 
    In this paper, we address the problem of statistical inference for parameters of interest in a target population by borrowing information from multi-source datasets, where the response variables are unobserved in the target population. This problem is motivated by a real-world example studying the relationship between ozone concentration ($\text{O}_3$) and meteorological variables that physically influence $\text{O}_3$ in a large spatial scale. Limited high-precision monitoring equipment causes O3 measurements available only at a small number of locations, while meteorological data are available on regular spatial grids. Therefore, it is important to leverage  $\text{O}_3$ measurements from multiple observational sites to estimate the model parameters at unmonitored locations. 
    
    %Existing transfer learning methods usually assume full data availability in both target and source domains \citep{Li2021TransLasso, Tian2023TransGlm} and thus are not directly applicable when the target response is completely missing.

Existing transfer learning methods primarily address high-dimensional supervised settings, where both responses and covariates are observed in both domains \citep{Li2021TransLasso, Tian2023TransGlm, Li2024Estimation}. \citet{Cai2024Triply} considers partially labeled target domains, leveraging both labeled and unlabeled data. However, these approaches are not applicable when response variables are completely unobserved in the target domain. Semi-supervised learning is a special case of covariate shift where labeled and unlabeled data share the same covariate distribution.
% It is worth noting that semi-supervised learning can be viewed as a special case of covariate shift, where the covariate distributions in the labeled and unlabeled data are assumed to be identical. 
\cite{song2023general} explored general loss functions for semi-supervised learning under a fixed-dimensional setting, while \cite{Deng2024optimal} considered high-dimensional linear models in the same context. 
However, standard semi-supervised learning methods do not account for distributional heterogeneity across sources. 

When responses are absent in the target domain, the problem shifts to estimating the source–target density ratio and the outcome regression. Doubly robust methods in the formulation of augmented inverse probability weighting (AIPW) have been proposed to deal with the missing data and causal inference problems \citep{bang2005doubly, funk2011doubly}.
%The challenge lies in combining multiple working models for robustness against model misspecification.
The AIPW estimator has been widely used in literature; see \cite{Tan2020Modelassited} for doubly robust confidence intervals with calibration and \cite{Zhou2024doubly} for doubly robust classification.
%\cite{Liu2023AugmentedSeminonpara} also introduced a doubly robust estimation procedure with semi-non-parametric nuisance models.
However, those approaches mostly focus on population mean or average treatment effects in single-source scenarios, which do not accommodate the challenges posed by multiple, heterogeneous sources. Classical AIPW methods use only one model for the outcome regression and one for the density ratio, limiting their capacity to integrate multiple working models. Moreover, the AIPW estimator may lead to nonconvex optimization for estimating equations and M-estimation, resulting in multiple local solutions and causing optimization challenges.

In this work, we study the problem of estimation and inference for parameters defined by estimating equations in the target population using data from multiple sources, where a high-dimensional set of covariates is observed but the response variables are completely missing in the target domain. We consider the high-dimensional covariate shift scenarios. Namely, the conditional distribution of the response given the high-dimensional covariates is the same between the source and target distributions, but the distributions of covariates could be different. 
%both the high-dimensional covariates and the response variables are available in multi-source datasets
We propose a statistical inference procedure for the target parameter that transfers the multi-source information to the target population and incorporates multiple working models for the density ratios and outcome regression. 

Specifically, we first develop a high-dimensional calibration weighting estimator that accommodates multiple working models by extending soft covariate balancing \citep{Zubizarreta2015} to high-dimensional settings. To eliminate the influence of high-dimensional nuisance parameters and obtain valid statistical inference, we develop a novel way of achieving Neyman orthogonality based on carefully designed calibration constraints, including orthogonality constraints and an outcome model projection constraint for debiasing. 
This leads to the high-dimensional debiased calibration (HDC) estimator. Different from the AIPW estimators that achieve double robustness by explicitly adding a debiasing term, the HDC estimator builds the debiasing step internally through the calibration constraints.
It attains asymptotic normality if either the true density ratio lies in the linear space spanned by the working models or the outcome regression model is correctly specified, ensuring multiply robust inference. 
Based on the asymptotic normality of the HDC estimators for all sources, we construct a specification test for the homogeneity of different sources, a necessary condition for the transferability of multiple sources to the target population.
%This extends prior works in \cite{Li2021TransLasso, Tian2023TransGlm}, where complete observations in the target population are required.
We further propose the multi-source high-dimensional debiased calibration (MHDC) estimator for efficiently combining multi-source data to estimate the target parameter.

We have three main contributions.
First, we propose a framework for accommodating multi-source data as well as multiple working models for density ratio and outcome regression with the multiply robust property. 
%ensuring valid asymptotic confidence intervals if either the true density ratio lies in the linear space spanned by the working models or the outcome regression model is correctly specified. Multiple high-dimensional pre-trained models for the response are also incorporated. 
This relaxes the requirement on the correct prior knowledge of the covariate shift mechanism or the outcome regression model. 
Second, the proposed estimator achieves Neyman orthogonality internally through calibration weights. It keeps a simple inverse probability weighting (IPW) form. For target parameters in M-estimation and estimating equation problems, the proposed method guarantees convex optimization and a unique solution. This circumvents potential non-convex optimization problems in AIPW estimators. To our knowledge, we propose the first estimator in the IPW form that can achieve multiply robust inference under high dimensionality.
%\textbf{For the outcome regression model, we only require correct specification at the true parameter value rather than for all possible values, making the model specification more attainable in practice.}
%This property is particularly valuable in practice, as relying on a single model would require stronger, and often unrealistic, prior knowledge of the covariate shift mechanism or the outcome regression.  
%Second, our estimating procedure \textbf{ provides} a hypothesis test for homogeneity of different sources, a necessary condition for the multiple sources to be transferred to the target population (Condition \ref{as:ignorability}).  
% which becomes crucial when informative source domains are unknown a prior. 
% This extends prior work in \cite{Li2021TransLasso} and \cite{Tian2023TransGlm} where complete observation in the target population was required.
Third, the proposed MHDC approach effectively combines multiple sources and attains efficiency improvements compared to combining estimators obtained by using each source separately. This highlights the advantage of the proposed estimator to incorporate data from multiple sources for multi-source learning.

This paper is structured as follows. Section \ref{sec:prelim} introduces the framework of multi-source data learning with missing responses, accompanied by a motivating example. 
% of a practical scientific problem. 
Section \ref{sec:HD-EL} develops the HDC estimating procedure. 
% with soft covariate balancing constraints and a pretrained conditional mean model. The augmented outcome regression is introduced to correct the bias for statistical inference. 
Section \ref{sec:test} proposes the specification test for the transferability condition of source data. Section \ref{sec:inference} generalizes the proposed method to multi-source settings. Section \ref{sec:asym} establishes the theoretical results for the proposed estimators.  
Sections \ref{sec:simulation} and \ref{sec: real data} provide the simulation studies and real data analysis on meteorological data, respectively.
% Section \ref{sec:simulation} 
% presents simulation studies that empirically validate the theoretical results and demonstrate a superior performance relative to existing methods .
% provides simulation studies, validating
% with emphasis on the performance of the debiasing procedure and multi-source learning. 
% The simulation results validate 
% the theory of the proposed estimator and shows the superiority over existing methods. 
% Section \ref{sec: real data} demonstrates practical utility through an analysis of ground ozone regression coefficients against multiple climatological variables. 
Section \ref{sec: conclusion} concludes the paper. All technical proofs and additional numerical results are relegated to the Supplementary Material (SM) \citep{SMtoMDEL}.

\setcounter{equation}{0}
\section{Setting and Examples}\label{sec:prelim}

Suppose that we are interested in establishing the relationship between the response variable $Y$ and target covariates $\bZ = (Z_{1}, \ldots, Z_{q})^{\T}$ under the target population $\mathbb{Q}$, where $Y$ is not observed on $\mathbb{Q}$. Let $\bW$ be a set of high-dimensional (HD) auxiliary covariates with dimension $(p - q)$, and $\bX = (\bZ^{\T}, \bW^{\T})^{\T} = (X_{1}, \ldots, X_{p})^{\T}$. Here, $X_j = Z_j$ for $j = 1, \ldots, q$. We assume $\bX$ is fully observed on $\mathbb{Q}$, the dimension $q$ of $\bZ$ is fixed, but the dimension of $\bX$ could be much larger than the sample size. 
Meanwhile, we observe multi-source data $\mathcal{S}_{\mathbb{F}, u} = \{(\bX_{u, i}, Y_{u, i}) \}_{i=1}^{n_{1 u}}$ of both $\bX$ and $Y$, independently and identically distributed (i.i.d.) from the $u$th source population with distribution $\mathbb{F}_u$ for $u = 1, \ldots, g$, where $\bX_{u, i} = (\bZ_{u, i}^{\T}, \bW_{u, i}^{\T})^{\T} = (X_{u, i1}, \ldots, X_{u, ip})^{\T}$.
We assume that $\lim_{n_{1u},n_0\to \infty} n_{1u}/n_0 = c_u \in (0,\infty) $ for all $u$.
Our goal is to estimate the unknown parameter $\bbeta = (\beta_1, \ldots, \beta_q)^{\T}$ that is the solution to the $q$-dimensional estimating equation 
\be 
{\E}_{\mathbb{Q}} \{\bS(Y, \bZ; \bbeta_{0})\} = \bzero
\label{eq:beta-0}\ee
on the target population $\mathbb{Q}$, where $\bS(Y, \bZ; \bbeta_{0}) = (S_1(Y, \bZ; \bbeta_{0}), \ldots, S_q(Y, \bZ; \bbeta_{0}))^{\T}$, $\bbeta_0 = (\beta_{0,1}, \ldots, \beta_{0,q})^{\T}$ is the true parameter value, and ${\E}_{\mathbb{Q}}(\cdot)$ denotes the expectation under the target population. To mainly focus on multi-source learning, we consider the just-identified case. The proposed method can be extended to over-identified estimating equations.

Let $\mathcal{S}_{\mathbb{Q}} = \{\bX_i\}_{i = 1}^{n_0}$ denote the unsupervised data i.i.d.\,from the target population, where $n_0$ is the sample size of $\mathcal{S}_{\mathbb{Q}}$, and $\bX_i = (\bZ_{i}^{\T}, \bW_{i}^{\T})^{\T} = (X_{i1}, \ldots, X_{ip})^{\T}$.
Let $\mathcal{S}_{\mathbb{F}} = \bigcup_{u=1}^g \mathcal{S}_{\mathbb{F}, u}$ be the union of the training data from all sources, and $n_1 = \sum_{u = 1}^{g} n_{1u}$. 
%If the conditional distribution of $Y \mid \bX$ is the same under the target and source populations, the information in the source data can be transferred to the target population. 
%In practice, $n_0$ could be larger than $n_1$ as supervised samples require annotation and labeling which are typically more expensive to collect.
%
% We mainly focus on three challenges in this work. First, as the transferability often holds with high-dimensional ancillary covariates $\bW$, it poses a difficulty to achieve valid inference of the target parameter $\bbeta_0$ even though it is in a low-dimensional space. Secondly, the estimating process leads to the hypothesis testing of transferability, i.e. Condition \ref{as:ignorability}, which becomes necessary when the informative source domains are unknown. Similar problems were discussed in  \cite{Li2021TransLasso,Tian2023TransGlm} where the response variable $Y$ was observed in the target domain. Third, we achieve an efficient estimation under the multi-site scenario, where the efficiency gain is attributed to the combination of different sources $\calS_{\mathbb{F},u}$ over linear combination of the estimators using each source separately.
%
Let $\mathbb{F} = \sum_{u=1}^g a_u \mathbb{F}_{u}$ be the overall source distribution of $\mathcal{S}_{\mathbb{F}}$, where $a_u = \lim n_{1u} / n_1$. 
%is the limiting sample size proportion of $n_{1u}$ to $n_1$.
Let $\mathbb{F}_{u, Y \mid \bX}$, $\mathbb{F}_{Y \mid \bX}$ and $\mathbb{Q}_{Y \mid \bX}$ denote the conditional distribution of $Y$ given $\bX$ under $\mathbb{F}_u$, $\mathbb{F}$ and $\mathbb{Q}$, respectively.
	Similarly, let $\mathbb{F}_{u, \bX}$, $\mathbb{F}_{\bX}$ and $\mathbb{Q}_{\bX}$ be the marginal distribution of $\bX$. 
	%We consider transfer learning under covariate shift, which refers to the case that $Y \mid \bX$ is distributed the same under $\mathbb{F}$ and $\mathbb{Q}$, but the marginal distributions of $\bX$ could be different. Namely, $\mathbb{F}_{Y \mid \bX} = \mathbb{Q}_{Y \mid \bX}$ and $\mathbb{F}_{\bX} \neq \mathbb{Q}_{\bX}$.
	%and $Y_i$ is the response variable of interest for $i = 1, \ldots, n_1$.
	Let $$w_u(\bx) = \frac{d \mathbb{Q}_{\bX}}{d \mathbb{F}_{u, \bX}} \mbox{ \ and \ } w(\bx) = \frac{d \mathbb{Q}_{\bX}}{d \mathbb{F}_{\bX}}$$ be the density ratios of $\mathbb{Q}_{\bX}$ to $\mathbb{F}_{u, \bX}$ and $\mathbb{F}_{\bX}$, respectively. 
    %If all sources share the same covariates, i.e. $\bX_{1} = \ldots = \bX_{g}$ , 
    Note that 
	\be
	\frac{1}{w(\bx)} = \sum_{u = 1}^g \frac{a_u}{w_u(\bx)}.
	\label{eq:LC-IPW}\ee
	To identify the parameter $\bbeta$ under the target population $\mathbb{Q}$ in (\ref{eq:beta-0}), we make the covariate shifting (transferability) assumption for the $u$th source population.
	%we assume the response variable and the indicator for supervised and unsupervised samples are independent given the covariates, as stated in the following ignorability condition.
	
	\begin{as}\label{as:ignorability}
	%The conditional distributions of $Y$ given $\bX$ are the same under $\mathbb{F}$ and $\mathbb{Q}$. And the 
	For the population $\mathbb{F}_{u}$, $\mathbb{Q}_{Y \mid \bX} = \mathbb{F}_{u, Y \mid \bX}$ and $c_0 < w_{u}(\bx) < c_0^{-1}$ for a small positive constant $c_0$ and any $\bx$ on the support of $\mathbb{F}_{u, \bX}$.
	\end{as}
 
    If Condition \ref{as:ignorability} holds for the $u$th source population, $\bbeta_0$ in (\ref{eq:beta-0}) can be identified via solving either
    ${\E}_{\mathbb{F}_u} \{w_u(\bX) \bS(Y, \bZ; \bbeta)\} = \bzero$ or 
	${\E}_{\mathbb{Q}_{\bX}} {\E}_{\mathbb{F}_{u, Y \mid \bX}} \{\bS(Y, \bZ; \bbeta)\} = \bzero$. Then, we can use the labeled data from this source to estimate $\bbeta_0$. 
    Here, we do not necessarily assume that the conditional distribution of $Y$ given $\bZ$ is the same. The transferability is valid when conditioning on both $\bZ$ and high-dimensional ancillary covariates $\bW$. The high-dimensional setting of $\bW$ makes Condition \ref{as:ignorability} more attainable. The marginal distributions of $\bX=(\bZ^\top, \bW^\top)^\top$ could be different between the source and target populations. 
    %where the former is based on density ratio weighting and the latter uses the outcome regression of $Y$ on $\bX$. 
	Semi-supervised learning assumes $\mathbb{Q} = \mathbb{F}_u$, which is a special case of the setting we consider. 
    In this paper, we consider three main problems regarding multi-source learning: how to obtain a robust inference of $\bbeta_0$ by combining multiple working outcome regression and density ratio models, how to test for the transferability of all sources, and how to construct an efficient estimator of $\bbeta_0$ by utilizing all transferable sources.

\medskip

{\bf Motivating example.}
Our motivating example is to investigate the relationship between ozone variation and key meteorological variables \citep{WangW2019AerosolsandOzone, Li2021radiative}. Ozone ($\text{O}_3$) plays an important role in the atmosphere, posing risks to human health, ecosystems, and vegetation. 
%Understanding its dynamics is crucial for air quality management and environmental protection.
The relationships between ozone and total solar radiation (TSR) and temperature (TEMP) provide insights into the mechanisms of ozone formation and degradation.
% , which is important for advancing air quality research, environmental pollution control, and policy formulation. 
However, the measurement of ozone requires specialized equipment and is typically missing in most data-scarce regions. To address this issue, we use the data from the source regions with complete observations to learn the target region without ozone measurement, where the source and target regions have a similar level of air pollution. In this example, the response $Y$ is $\text{O}_3$ concentration, the target covariates $\bZ$ include TSR, TEMP and the lag terms, and the auxiliary covariates $\bW$ consist of a high-dimensional set of environmental factors, including meteorological variables and their lag terms, which may influence ozone levels and can be collected over all regions by remote sensing measurements.

To study the linear effect of $\bZ$ on $\text{O}_3$ concentration, we consider the linear working model $Y = \bZ^{\T} \bbeta + \epsilon$, where the target parameter $\bbeta_0 = {\E}_{\mathbb{Q}}^{-1} (\bZ \bZ^{\T}) {\E}_{\mathbb{Q}}(\bZ Y)$ corresponds to ${\E}_{\mathbb{Q}}(\bZ \epsilon) = \bzero$, and the estimating equation is $\bS(Y, \bZ; \bbeta) = (Y - \bZ^{\T} \bbeta) \bZ$. Additionally, the partial correlation between $\text{O}_3$ and TSR can be estimated via linear node-wise regression \citep{QZ2021}.
%which provides a useful measure for linear association between ozone and the specific variable, say total solar radiation(TSR), after controlling the linear effects of all other variables.
% This example is related to statistical inference for partial correlation via nodewise regression \citep{qiu2020estimating, QZ2021}. Under non-Gaussian distributions, partial correlation is not equivalent to conditional correlation since ${\E}(\bY \mid \bZ) \neq \bZ; \bbeta$. But, it is still a useful measure for conditional dependence, which describes the linear association between two variables after controlling the linear effects of all other variables. 
If researchers are interested in whether ozone concentrations exceed a specific level $\tau$, generalized linear models (GLMs) can be applied and the target parameter $\bbeta_0$ is the solution to the estimating equation $ {\E}_{\mathbb{Q}}[ \{ Y^{(\tau)} - b(\bZ; \bbeta)\} \bZ] = \bzero$, where $Y^{(\tau)} = \mathbb{I}(Y > \tau)$ and $b(\bZ; \bbeta)$ is the working model for ${\E}_{\mathbb{Q}}( Y^{(\tau)} | \bZ)$. Note that the true conditional expectations ${\E}_{\mathbb{Q}}(Y | \bZ)$ and ${\E}_{\mathbb{Q}}(Y^{(\tau)} | \bZ)$ may not equal $\bZ^{\T} \bbeta$ or $b(\bZ; \bbeta)$, and $\bbeta_{0}$ is the coefficient for the best prediction of the target response given $\bZ$ under the working model.

\setcounter{equation}{0}
\section{HD Debiased Calibration Weighting} \label{sec:HD-EL}

In this section, we first consider single-source learning and develop high-dimensional calibration weighting for estimating $\bbeta$, which can combine multiple density ratio (DR) and outcome regression (OR) models. 
Let $w_{u, k}(\bx, \bgamma_{u, k})$ be the $k$th working model for the density ratio $w_u(\bx)$, where $k = 1, \ldots, v_u$. 
We say the $k$th model is correctly specified if 
\be\label{eq:HD-DR}
w_u(\bx) = w_{u, k}(\bx, \bgamma_{u, k}^{(0)})
\ee
for all $\bx$ and some coefficient $\bgamma_{u, k}^{(0)}$.
Let $m(\bx, \bxi)$ be the working model for the outcome regression ${\E}_{\mathbb{Q}}(Y | \bX = \bx)$. For simplicity of presentation, we only consider one OR model. Let $ \widehat{\bgamma}_{u, k}$ and $\widehat{\bxi}_u$ be the estimates of $\bgamma_{u, k}$ and $\bxi$ using the data from $\mathcal{S}_{\mathbb{F}_u}$ and $\mathcal{S}_{\mathbb{Q}}$. 
%Then, $w_u(\bx, \widehat{\bgamma}_u)$ and $m(\bx, \widehat{\bxi}_u)$ are the estimates of $w_u(\bx, \bgamma_u)$ and $m(\bx, \bxi)$. 

Suppose the estimating equation can be decomposed as 
\begin{equation}
\bS(Y,\bZ;\bbeta) = \bS_{a}(Y,\bZ;\bbeta) + \bS_{b}(\bZ,\bbeta),
\label{eq:3-1}
\end{equation}
where the second part $\bS_{b}(\bZ,\bbeta)$ is free of $Y$. This decomposition is satisfied for the linear model equation $\bS(Y, \bZ; \bbeta) = (Y - \bZ^{\T} \bbeta) \bZ$ with $\bS_{a}(Y,\bZ;\bbeta) = Y \bZ$ and $\bS_{b}(\bZ, \bbeta) = (\bZ^{\T}\bbeta) \bZ$, and the GLM equation $\bS(Y, \bZ; \bbeta) = \{ Y - b(\bZ; \bbeta)\} \bZ$ with $\bS_{a}(Y, \bZ;\bbeta) = Y \bZ$ and $\bS_{b}(\bZ; \bbeta) = b(\bZ; \bbeta) \bZ$. 
If $\bS(Y,\bZ;\bbeta)$ cannot be decomposed as in (\ref{eq:3-1}), we may take $\bS_{b}(\bZ,\bbeta) = \bzero$. Let $S_{aj}(Y,\bZ;\bbeta)$ and $S_{bj}(\bZ,\bbeta)$ be the $j$th coordinate of $\bS_{a}(Y,\bZ;\bbeta)$ and $\bS_{b}(\bZ, \bbeta)$, respectively. 

We view the OR model $m(\bx, \widehat{\bxi}_u)$ as a pre-trained model for the response and use an additional set of basis functions $\bb(\bx) = (b_1(\bx), \ldots, b_{m}(\bx))^{\T}$ to model ${\E}_{\mathbb{Q}} \{\bS_{a}(Y, \bZ; \bbeta) | \bX\}$, where $b_1(\bx) = 1$ and the dimension $m$ could be much larger than the sample sizes. 
%The basis functions $\bb(\bx)$ may exhibit dependence on the parameter $\bbeta$, however, for the sake of notational simplicity, we suppress this dependence in our presentation. 
For a given function $\widetilde{S}_{aj}(Y,\bZ)$, the working OR model for 
${\E}_{\mathbb{Q}} \{{S}_{aj}(Y, \bZ; \bbeta) | \bX\}$ is
% However, as ${\E}_{\mathbb{F}_{u}} \{L(Y, \bZ; \bbeta) \mid \bX\} \neq L({\E}_{\mathbb{F}_{u}}(Y \mid \bX), \bZ; \bbeta)$ in general, we consider to use a set of basis functions $\bb(\bx) = (b_1(\bx), \ldots, b_{m}(\bx))^{\T}$ to model the difference between ${\E}_{\mathbb{F}_{u}} \{L(Y, \bZ; \bbeta) \mid \bX\}$ and $L({\E}_{\mathbb{F}_{u}}(Y \mid \bX), \bZ; \bbeta)$. Our working OR model for the $u$th source is
\be
{\E}_{\mathbb{Q}} \{{S}_{aj}(Y, \bZ; \bbeta) \mid \bX = \bx\}  \in \operatorname{span}\{\widetilde{S}_{aj}(m(\bx, \bxi), \bZ), b_1(\bx), \ldots, b_{m}(\bx)\}, \label{eq:HD-OR}
% {\E}_{\mathbb{F}_{u}} \{L(Y, \bZ; \bbeta) \mid \bX = \bx\} \in \operatorname{span}\{L(m_u(\bx, \bxi_u), \bz; \bbeta), b_1(\bx), \ldots, b_{m}(\bx)\}, \label{eq:HD-OR}
\ee
% where $\widetilde{S}_j(Y, \bZ; \bbeta) = S_j(Y, \bZ; \bbeta) - \widetilde{S}_{j2}(\bZ, \bbeta)$.
%\textcolor{red}{Working outcome regression model is 
%$${\E}_{\mathbb{F}_{u}} \{L(Y, \bZ; \bbeta) \mid \bX = \bx\} \in \operatorname{span}\{  b_1(\bx), \ldots, b_{m}(\bx)\},$$}
where $\widetilde{S}_{aj}(Y,\bZ)$ is constructed to capture the relationship between $Y$ and $\bZ$ in $S_{aj}(Y,\bZ;\bbeta)$.
For the linear model and GLM, we can take $\widetilde{S}_{aj}(Y, \bZ) = S_{aj}(Y,\bZ;\bbeta) = Y Z_j$. 
%Here, $m(\bx, \bxi_u)$ is a working model of ${\E}_{\mathbb{Q}}(Y | \bX)$ for pre-trained calibration.
As ${S}_{bj}(\bZ, \bbeta)$ is a function of $\bZ$ and $\bbeta$ only, there is no need to estimate its mean on $\mathbb{Q}$ by outcome regression or density ratio weighting.
We remove it from $S_j(Y, \bZ; \bbeta)$ so that the OR model in (\ref{eq:HD-OR}) only concerns the part of the estimating equations with $Y$.

\subsection{Single source calibration weighting}
\label{sec: HD-EL-estimation}

Since $ {\E}_{\mathbb{F}_u} \{w_u(\bX) a(\bX)\} = {\E}_{\mathbb{Q}} \{a(\bX)\}$ for any function $a(\bX)$ of $\bX$, the calibration estimation (also called covariate balancing estimation) of the weights $\{w_u(\bX_i)\}$ can be obtained by solving the sample version of this calibration equation using a fixed number of $a(\bX)$ \citep{hainmueller2012entropy, Han2014}. 
However, directly solving the calibration weights for high-dimensional calibration functions $a(\bx)$ is not possible, where exact covariate balancing would not have a solution for the weights.
To tackle this problem, we propose high-dimensional weighting with soft covariate balancing (SCB) constraints.
%We extend the classical covariate balancing approach to a high-dimensional set of calibration functions in (\ref{eq:HD-OR})  via an EL formulation with soft covariate balancing (SCB). 
Let $\widetilde{\bS}_{a}(m(\bx, \bxi), \bZ) = (\widetilde{S}_{a1}(m(\bx, \bxi), \bZ), \ldots, \widetilde{S}_{aq}(m(\bx, \bxi), \bZ))^{\T}$, $\bar{\bS}_{a,u} = n_0^{-1} \sum_{i \in \mathcal{S}_{\mathbb{Q}}} \widetilde{\bS}_{a}(m(\bX_i, \widehat{\bxi}_u), \bZ_i)$, $\bar{\bb}_0 = (\bar{b}_{0, 1}, \ldots, \bar{b}_{0, m})^{\T} = n_0^{-1} \sum_{i \in \mathcal{S}_{\mathbb{Q}}} \bb(\bX_i)$, and $\bb_{c}(\bX_i) \allowbreak  = (b_{c,1}(\bX_i), \ldots, b_{c,m}(\bX_i))^{\T} = \bb(\bX_i) - \bar{\bb}_0$. 
%where $\widetilde{\bbeta}$ is an initial estimate of $\bbeta$.
% Let $\bar{L}_{u, 0} = n_0^{-1} \sum_{i \in \mathcal{S}_{\mathbb{Q}}} L(m_u(\bX_i, \widehat{\bxi}_u), \bZ_i; \bbeta)$, $\bar{\bb}_0 = (\bar{b}_{0, 1}, \ldots, \bar{b}_{0, m})^{\T} = n_0^{-1} \sum_{i \in \mathcal{S}_{\mathbb{Q}}} \bb(\bX_i)$ and $\bb_{c}(\bX_i) = (b_{c,1}(\bX_i), \ldots, b_{c,m}(\bX_i))^{\T} = \bb(\bX_i) - \bar{\bb}_0$, where $\widetilde{\bbeta}$ is an initial estimate of $\bbeta$.
%
%$$\bigg|\frac{1}{n_1}\sum_{i = 1}^{n} D_i w(\bX_i) b(\bX_i) - \frac{1}{n_0}\sum_{i = 1}^{n} (1 - D_i) b(\bX_i)\bigg| \leq c_n$$
%for a positive sequence $c_n \to 0$ as $n \to \infty$.
%Given the initial estimate $\widetilde{\bbeta}$, 
We solve the following high-dimensional empirical likelihood (EL) with both equality and inequality constraints for calibration weighting: 
% \footnote{YQ: I add subscript $u$ to $\widehat{p}_i(\bbeta)$. I basically add the subscript $u$ to everywhere. In the proof, you may say ``For simplicity of the notation and presentation, we drop the subscript $u$ whenever there is no confusion.''}
% \bea
% \{\widehat{p}_i(\bbeta): i \in \mathcal{S}_{\mathbb{F}}\} &=& \underset{ \{ p_i: i \in \mathcal{S}_{\mathbb{F}} \} }{\operatorname{argmax}} \sum_{i \in \mathcal{S}_{\mathbb{F}}} \log(p_i), \mbox{ \ subject to \ } p_i \geq 0, \ \sum_{i \in \mathcal{S}_{\mathbb{F}}} p_i = 1, \label{eq:EL-observe} \\
% && \sum_{i \in \mathcal{S}_{\mathbb{F}}} p_i \bigg\{ \frac{1}{1 - \pi_{k}(\bX_i, \widehat{\bgamma}_{ k})} - \frac{n_{1} + n_0}{n_0} \bigg\} = 0 \mbox{ \ for $k \in [v]$,} \ \ \ \ \ \ \  \label{eq:EL-IBC} \\
% && \sum_{i \in \mathcal{S}_{\mathbb{F}}} p_i \{ L(m(\bX_i, \widehat{\bxi}), \bZ_i; \bbeta) - \bar{L}_{0} \} = 0 , \label{eq:EL-CB-exact} \\
% && \bigg| \sum_{i \in \mathcal{S}_{\mathbb{F}}} p_i b_{c,j}(\bX_i) \bigg| \leq \omega_{\ps} %\sqrt{\frac{\log m}{n}} 
% \mbox{ \ for $j \in [m]$}, \label{eq:EL-CB}
% \eea
\bea
\{\widehat{p}_{u,i}: i \in \mathcal{S}_{\mathbb{F}_u}\} &=& \underset{ \{ p_i: i \in \mathcal{S}_{\mathbb{F}_u} \} }{\operatorname{argmax}} \sum_{i \in \mathcal{S}_{\mathbb{F}_u}} \log(p_i), \mbox{ \ subject to \ } p_i \geq 0, \ \sum_{i \in \mathcal{S}_{\mathbb{F}_u}} p_i = 1, \label{eq:EL-observe} \\
&& \sum_{i \in \mathcal{S}_{\mathbb{F}_u}} p_i w_{u, k}^{-1}(\bX_i, \widehat{\bgamma}_{u, k}) = 1 \mbox{ \ for $k = 1, \ldots, v_u$,} \label{eq:EL-IBC} \\
&& \sum_{i \in \mathcal{S}_{\mathbb{F}_u}} p_i \widetilde{\bS}_{a}(m(\bX_i, \widehat{\bxi}_u), \bZ_i) = \bar{\bS}_{a, u}, \label{eq:EL-CB-exact} \\
&& \bigg| \sum_{i \in \mathcal{S}_{\mathbb{F}_u}} p_i b_{c,j}(\bX_i) \bigg| \leq \omega_{\ps} \mbox{ \ for $j = 1, \ldots, m$}, \label{eq:EL-CB}
\eea
where $\omega_{\ps} > 0$ is the tolerance parameter for non-exact covariate balancing of $\bb(\bX_i)$.
Here, (\ref{eq:EL-IBC}) is the internal bias correction (IBC) constraint which calibrates the weights $\{\widehat{p}_{u,i}(\bbeta)\}$ to the estimates $w_{u, k}(\bX_i, \widehat{\bgamma}_{u, k})$ from the $k$th working DR model; (\ref{eq:EL-CB-exact}) is the pre-trained model calibration (PMC) constraint which incorporates the pre-trained model $m(\bX_i, \widehat{\bxi}_u)$ into the weights; (\ref{eq:EL-CB}) is the soft covariate balancing constraints \citep{Zubizarreta2015} for high-dimensional basis functions, which is the key to extending the fixed-dimensional covariate balancing to high-dimensional settings. 
It is shown in the SM that $(n_{1u}\widetilde{p}_{u, i})^{2} \to w^2_u(\bX_i)$ as $n, p \to \infty$ if one of the working density ratio models $\{w_{u, k}(\bx, \bgamma_{u, k})\}$ is correctly specified.

%The covariate information of the unsupervised sample $\mathcal{S}_0$ is used in the SCB constraints.
%
%\cite{Han2014} considered a multiple robust estimator via fixed-dimensional covariate balancing for a missing data problem. Their IBC condition is $\sum_{i = 1}^{n} D_i p_i \{\pi_k(\bX_i, \widehat{\bgamma}_k) - n_1 / n\} = 0$ and their covariate balancing condition matches to the average of $\bb(\bX_i)$ over the whole sample of $\mathcal{S}_1$ and $\mathcal{S}_0$. Since our target is the parameters of the distribution $\mathbb{Q}$ of the unsupervised sample $\mathcal{S}_0$, our IBC condition in (\ref{eq:EL-IBC}) is built on the inverse PS $\{1 - \pi_{lk}(\bX_i, \widehat{\bgamma}_{lk})\}^{-1}$ instead of $\pi_{lk}(\bX_i, \widehat{\bgamma}_{lk})$. 
%The SCB constraint is the key to extending the fixed-dimensional covariate balancing to high-dimensional settings, where the weighted mean $\sum_{i \in \mathbb{F}} p_i b_{j}(\bX_i)$ of the balancing function under the source sample may not be exactly equal to its mean under the target unsupervised sample, and $\omega_{\ps}$ is the tuning parameter regulates the difference.

%Let $\widetilde{\bS}(Y, \bZ; \bbeta) = (\widetilde{S}_1(Y, \bZ; \bbeta), \ldots, \widetilde{S}_q(Y, \bZ; \bbeta))^{\T}$ and $\widetilde{\bS}_{2}(\bZ, \bbeta) = (\widetilde{S}_{12}(\bZ, \bbeta), \ldots, \widetilde{S}_{q2}(\bZ, \bbeta))^{\T}$. 
By (\ref{eq:3-1}), using the calibration weights $\{\widehat{p}_{u,i}\}$ from the $u$th source sample, the high-dimensional calibration weighted estimate $\widehat{\bbeta}_{\HC, u}$ of $\bbeta$ solves the equation 
\be
\sum_{i \in S_{\mathbb{F}_u}} \widehat{p}_{u,i} \bS_{a}(Y_i, \bZ_i; \bbeta) + \frac{1}{n_0} \sum_{i \in S_{\mathbb{Q}}} \bS_{b}(\bZ_i, \bbeta) = \bzero.
\label{eq:obj}\ee
The first term of (\ref{eq:obj}) estimates ${\E}_{\mathbb{Q}} \{\bS_{a}(Y, \bZ; \bbeta) \}$ from the source sample and the second term estimates ${\E}_{\mathbb{Q}} \{\bS_{b}( \bZ; \bbeta) \}$
from the target sample directly.
	
Let $\widehat{\bw}_{u,i} = (w_{u, 1}(\bX_i, \widehat{\bgamma}_{u, 1}), \ldots, w_{u, v_u}(\bX_i, \widehat{\bgamma}_{u, v_u}))^{\T}$ be the set of all estimated DR models and $\mathbf{1}_{v_u}^{\T}$ be a $v_u$-dimensional vector of ones. 
% Let $\widehat{\bpi}_i = (\pi_{1, 1}(\bX_i, \widehat{\bgamma}_{1, 1}), \ldots, \pi_{1, v_1}(\bX_i, \widehat{\bgamma}_{1, v_1}), \ldots, \pi_{G, v_G}(\bX_i, \widehat{\bgamma}_{G, v_G}))^{\T}$ be the set of all estimated PS models from all sources and $\mathbf{c}_{n} = (\frac{n_{11}+n_0}{n_0} \mathbf{1}_{v_1}^{\T}, \ldots, \frac{n_{1G}+n_0}{n_0} \mathbf{1}_{v_G}^{\T})^{\T}$ be a $v$ dimensional vector, where $v = \sum_{u=1}^G v_u$ and $\mathbf{1}_{v_u}$ denotes a $v_u$ dimensional vector of ones. 
% $\mathbf{1}_v = (1, \ldots, 1)^{\T}$ be a $v = \sum_{l=1}^L v_l$-dimensional vector of 1.
% Let $(\mathbf{1}_v - \widehat{\bpi}_i)^{-1}$ denote the entry-wise reciprocal operator. 
From the Lagrange form of the constrained optimization problem in (\ref{eq:EL-observe}), the calibration weights $\{\widehat{p}_{u,i}: i \in \mathcal{S}_{\mathbb{F}_u}\}$ take the form
\be
\widehat{p}_{u,i} = n_{1u}^{-1} \big[ \widehat{\lambda}_{u,0} + \widehat{\blambda}_{u,1}^{\T} \{\widehat{\bw}_{u, i}^{-1}  - \mathbf{1}_{v_u}\} + \widehat{\blambda}_{u,2}^{\T} \bb_c(\bX_i) + \widehat{\blambda}_{u,3}^{\T} \{ \widetilde{\bS}_{a}(m(\bX_i, \widehat{\bxi}_u), \bZ_i) - \bar{\bS}_{a, u} \} \big]^{-1},
\label{eq:EL-weights}\ee
where $\widehat{\lambda}_{u,0}$, $\widehat{\blambda}_{u,1}  = (\widehat{\lambda}_{u,1 1}, \ldots, \widehat{\lambda}_{u,1v_u})^{\T}$, $\widehat{\blambda}_{u,2} = (\widehat{\lambda}_{u,21}, \ldots, \widehat{\lambda}_{u,2m})^{\T}$ and $\widehat{\blambda}_{u,3}$ are the Lagrange multipliers such that $\{\widehat{p}_{u,i}: i \in \mathcal{S}_{\mathbb{F}_u}\}$ satisfy the constraints in (\ref{eq:EL-IBC}), (\ref{eq:EL-CB-exact}) and (\ref{eq:EL-CB}).
    %where $\widehat{\lambda}_{0}$, $\widehat{\blambda}_{1} = (\widehat{\blambda}_{1 1}^{\T}, \ldots, \widehat{\blambda}_{1 G}^{\T})^{\T}$, $\widehat{\blambda}_{1 u} = (\widehat{\lambda}_{1 u 1}, \ldots, \widehat{\lambda}_{1 u v_u})^{\T}$, $\widehat{\blambda}_{2} = (\widehat{\lambda}_{21}, \ldots, \widehat{\lambda}_{2m})^{\T}$ and $\widehat{\blambda}_{3} = (\widehat{\lambda}_{3 1}, \ldots, \widehat{\lambda}_{3 G})^{\T}$ are the Lagrange multipliers that satisfy the constraints in (\ref{eq:EL-IBC}), (\ref{eq:EL-CB-exact}) and (\ref{eq:EL-CB}).
	%\bea
	%&& \frac{1}{n_1} \sum_{i = 1}^{n} \frac{D_i}{\widehat{\lambda}_0 + \widehat{\blambda}_{1}^{\T} \{(\mathbf{1}_v - \widehat{\bpi}_i)^{-1}  - \mathbf{c}_n\} + \widehat{\blambda}_{2}^{\T} \bb_c(\bX_i)} = 1 \nn \\
	%&& \frac{1}{n_1} \sum_{i = 1}^{n} \frac{D_i [\{1 - \pi_{lk}(\bX_i, \widehat{\bgamma}_{lk})\}^{-1} - \frac{n_l+n_0}{n_0}] }{\widehat{\lambda}_0 + \widehat{\blambda}_{1}^{\T} \{(\mathbf{1}_v - \widehat{\bpi}_i)^{-1}  - \mathbf{c}_n\} + \widehat{\blambda}_{2}^{\T} \bb_c(\bX_i)} = 0 \label{eq:EL-lambda-equation-1}
	%\eea
	%for $l \in [G], k \in [v_l]$ and
	%\bea
	%\bigg| \frac{1}{n_1} \sum_{i = 1}^{n} \frac{D_i b_{c,j}(\bX_i)}{\widehat{\lambda}_0 + \widehat{\blambda}_{1}^{\T} \{(\mathbf{1}_v - \widehat{\bpi}_i)^{-1}  -  \mathbf{c}_n \} + \widehat{\blambda}_{2}^{\T} \bb_c(\bX_i)} \bigg| \leq \omega_{\ps} \sqrt{(\log m) / n} \nn
	%\eea
	%for $j = 1, \ldots, m$.
Let $|\cdot|_1$ denote the $\ell_1$ norm. By the Karush–Kuhn–Tucker (KKT) condition, $\widehat{\lambda}_{u,0}$, $\widehat{\blambda}_{u,1}$, $\widehat{\blambda}_{u,2}$, and $\widehat{\blambda}_{u,3}$ can be obtained by solving the penalized unconstrained optimization problem:
\bea
&\underset{\lambda_{0}, \blambda_{1}, \blambda_{2}, \blambda_{3}}{\operatorname{argmin}} & \frac{1}{n_{1 u}} \sum_{i \in \mathcal{S}_{\mathbb{F}_u}} -\log\big[ \lambda_0 + \blambda_{1}^{\T} \{\widehat{\bw}_{u,i}^{-1}  - \mathbf{1}_{v_u}\} + \blambda_{2}^{\T} \bb_c(\bX_i) \nn \\
&& \ \ \ \ \ \ \ \ \ \ \ \ \ \ \ \ \ \ \ 
+ \blambda_{3}^{\T} \{ \widetilde{\bS}_{a}(m(\bX_i, \widehat{\bxi}_u), \bZ_i) - \bar{\bS}_{a, u} \} \big] + \omega_{\ps} |\blambda_{2}|_1 + \lambda_0.
\label{eq:EL-pi-1}
\eea

\subsection{Debiased calibration weighting}
\label{sec: single source inference}

Due to the non-exact covariate balancing in (\ref{eq:EL-CB}) and the regularized estimates $\widehat{\bgamma}_{u,k}$ and $\widehat{\blambda}_{u,k}$, the convergence rate of $\widehat{\bbeta}_{\HC, u}$ to $\bbeta_0$ is slower than the order $n_{0}^{-1/2}$. The classical statistical inference of Z-estimation is not applicable for $\widehat{\bbeta}_{\HC, u}$ in (\ref{eq:obj}).  
In this part, based on $\widehat{\bbeta}_{\HC, u}$ and $\widehat{p}_{u,i}$ from the initial calibration weighting, we propose a Neyman orthogonal estimating equation for $\bbeta$, which leads to a high-dimensional debiased calibration estimator $\widehat{\bbeta}_{\HDC, u}$ and a statistical inference procedure for $\bbeta_0$.

Let $w_{u, k}^{(1)}(\bX, \bgamma_{u, k}) = \frac{\partial}{\partial \bgamma_{u, k}} w_{u, k}(\bX, \bgamma_{u, k})$ be the first-order derivative of the $k$th working DR model. 
%If $\pi_{ k}(\bX, \bgamma_{ k})$ is from the family of logistic regression models, we have $\pi_{k}^{(1)}(\bX, \bgamma_{ k}) = \pi_{ k}(\bX, \bgamma_{ k}) \{1 - \pi_{ k}(\bX, \bgamma_{ k})\}$. 
Let $\widetilde{\bS}_a^{(1)}(\bX; \bxi_u) = (\frac{\partial}{\partial \bxi_u}\widetilde{S}_{a1}^{\T}(m(\bX, \bxi_u), \bZ),\ldots, \frac{\partial}{\partial \bxi_u}\widetilde{S}_{aq}^{\T}(m(\bX, \bxi_u), \bZ))^{\T} $ be the derivatives of the predicted score function with respect to $\bxi_u$. For a positive integer $c$, let $[c] = \{1, \ldots, c\}$. In addition to the functions $\bb(\bx)$ in (\ref{eq:HD-OR}), we consider an augmented set
% \be
% \widehat{\bfd}(\bx) = \bigg(\bb^{\T}(\bx), \bS_{j}^{(1) \T}(\bx; \widehat{\bxi}, \bbeta), \frac{\pi_{k}^{(1) \T}(\bx, \widehat{\bgamma}_{k})}{\{1 - \pi_{k}(\bx, \widehat{\bgamma}_{k})\}^{2}} :  k \in [v], j \in [q] \bigg)^{\T}
% \label{eq:basis-HD}\ee 
\be
\widehat{\bfd}_u(\bx) = \big(\bb^{\T}(\bx), \widetilde{\bS}_a^{(1) \T}(\bx; \widehat{\bxi}_u), {w_{u, k}^{(1)\T}(\bx,\widehat{\bgamma}_{u, k})}{ w_{u, k}^{-2}(\bx,\widehat{\bgamma}_{u, k})}  :  k \in [v_u] \big)^{\T}
\label{eq:basis-HD}\ee 
of functions for covariate balancing and outcome regression, that include the first-order derivatives of the predicted score and all working DR models.
%which includes the derivatives of all working PS models and the regression model $m(\bX, \bxi)$ of ${\E}_{\mathbb{F}}(Y \mid \bX)$.
	%where the polynomials $\{\pi_{k}^{l}(\bX, \widehat{\bgamma}_k): 0 \leq l \leq u \}$ are used to approximate the functional and derivatives of $\pi_{k}(\bX, \bgamma_k)$. Here, $u$ can be chosen at the order $\log(n)$ as $\pi_{k}^{l}(\bX, \widehat{\bgamma}_k)$ decays exponentially fast to zero with the increase of $l$. Note that restricting $l = 0$ corresponds to the original basis function $\bb(\bX)$. 
Let $\bar{\bfd}_{u, 0} = n_0^{-1} \sum_{i \in \mathcal{S}_{\mathbb{Q}}} \widehat{\bfd}_u(\bX_i)$ and $\widehat{\bfd}_{u, c}(\bX_i) = (\widehat{d}_{u, c, 1}(\bX_i), \ldots, \widehat{d}_{u, c, \widetilde{m}_u}(\bX_i))^{\T} = \widehat{\bfd}_u(\bX_i) - \bar{\bfd}_{u, 0}$, where $\widetilde{m}_u$ is the dimension of $\widehat{\bfd}_u(\bX)$. 
%Let $|\cdot|_{\infty}$ denote the element-wise maximum norm of a vector or matrix.
%For the high-dimensional scenario where the number of covariates is large, exact covariate balancing in (\ref{eq:EL-CB-fixedD}) is not feasible. We need to develop a new method for high-dimensional semi-supervised learning. Recall that $\widehat{\bgamma}_k$ is the estimate of $\bgamma_k$ and $\pi_{k}^{(1)}(\bX, \bgamma_k)$ is the first order derivative of the $k$th working model $\pi_k(\bX, \bgamma_k)$. 
%Given an initial estimate $\widetilde{\bbeta}$, 

For the $u$th source data, consider the regularized weighted regression with weights $\widehat{p}_{u,i}$:
\bea
(\widetilde{\btheta}_{u,j,1}^{\T}, \widetilde{\btheta}_{u,j,2}^{\T}, \widetilde{\btheta}_{u,j,3}^{\T}) & = 
& \underset{\btheta_{j,1}, \btheta_{j,2}, \btheta_{j,3}}{\operatorname{argmin}} \frac{1}{2n_{1 u}} \sum_{i \in \mathcal{S}_{\mathbb{F}_u}} (n_{1 u}\widehat{p}_{u, i})^{2} \{ {S}_{aj}(Y, \bZ; \widehat{\bbeta}^{\ini}) - \btheta_{j,1}^{\T} \widehat{\bfd}_u(\bX_i) \nn  \\
&& - \btheta_{j,2}^{\T} \widehat{\ba}_{u,1i} - \btheta_{j,3}^{\T} \widehat{\ba}_{u,2i} \}^2 + \omega_{\oc} 
%\sqrt{(\log \widetilde{m}) / n} 
(|\btheta_{j,1}|_1 + |\btheta_{j,2}|_1 + |\btheta_{j,3}|_1) \label{eq:EL-outcome}
\eea
for $j = 1, \ldots, q$, where we can take $\widehat{\bbeta}^{\ini} = \widehat{\bbeta}_{\HC, u}$ as the initial estimate of $\bbeta_0$, $\widehat{p}_{u,i}$ is the initial calibration weights from (\ref{eq:EL-weights}), 
%that satisfies the convergence rate specified in Theorem \ref{thm: estimating eqs convergence}, 
$\omega_{\oc}$ is a penalty parameter for the augmented outcome regression, $\widehat{\ba}_{u,1i} = \widetilde{\bS}_a(m(\bX_i, \widehat{\bxi}_u), \bZ_i)$ and $\widehat{\ba}_{u,2i} = \widehat{\bw}_{u, i}^{-1} = (w_{u, 1}^{-1}(\bX_i, \widehat{\bgamma}_{u, 1}), \ldots, w_{u, v_u}^{-1}(\bX_i, \widehat{\bgamma}_{u, v_u}))^{\T}$. 

Let $\widehat{\bU}_{u,i} =(\widehat{\bfd}_u(\bX_i)^{\T},\widehat{\ba}_{u,1i}^{\T}, \widehat{\ba}_{u,2i}^{\T})^{\T}$ denote the regressor of the augmented outcome regression, and $\widetilde{\btheta}_{u,j} = (\widetilde{\btheta}_{u,j,1}^{\T}, \widetilde{\btheta}_{u,j,2}^{\T}, \widetilde{\btheta}_{u,j,3}^{\T})^{\T}$ denote the estimated coefficients for the $u$th source data.
We obtain the calibration weights by solving the constraint optimization problem:
\bea
\{\widetilde{p}_{u, i}: i \in \mathcal{S}_{\mathbb{F}_u}\} &=& \underset{ \{p_i: i \in \mathcal{S}_{\mathbb{F}_u}\} }{\operatorname{argmax}} \sum_{i \in \mathcal{S}_{\mathbb{F}_u}} \log(p_i), \mbox{ \ subject to \ } p_i \geq 0, \ \sum_{i \in \mathcal{S}_{\mathbb{F}_u}} p_i = 1, \label{eq:EL-observe-inference} \\
&& \sum_{i\in S_{\mathbb{F}_u}} p_i \widehat{\bU}_{u,i}^{\T}\widetilde{\btheta}_{u,j} = \frac{1}{n_0}\sum_{i\in S_{\mathbb{Q}}} \widehat{\bU}_{u,i}^{\T}\widetilde{\btheta}_{u,j}  \mbox{ \ for $j = 1, \ldots, q$,} 
\label{eq:EL-CB-projection} \\
&& \mbox{Equations (\ref{eq:EL-IBC}), (\ref{eq:EL-CB-exact}) and } \ 
\bigg| \sum_{i \in \mathcal{S}_{\mathbb{F}_u}} p_i \widehat{\bfd}_{u, c}(\bX_i) \bigg|_{\infty} \leq \omega_{\ps}, \ \ \ \label{eq:EL-CB-HD-singlesource} 
\eea
where $|\cdot|_{\infty}$ denotes the element-wise maximum norm of vector or matrix. Compared to the original optimization problem in (\ref{eq:EL-observe})-(\ref{eq:EL-CB}), we introduce an additional constraint (\ref{eq:EL-CB-projection}) to balance the estimated projection $\widehat{\bU}_{u}^{\T}\widetilde{\btheta}_j$ of $S_{aj}(Y,\bZ;\bbeta)$ onto the augmented basis, which is called the projection calibration (PC) constraint. It plays a central role in incorporating the outcome model and ensuring the multiply robust property of the estimator.
The calibration constraints on the derivatives of $\widetilde{\bS}_a(\bX; \bxi_u)$ and $w_{u, k}^{-1}(\bX, \bgamma_{u, k})$ in (\ref{eq:EL-CB-HD-singlesource}) are called soft orthogonality (SO) constraints which removes the first order effects of $\widehat{\bxi}_u$ and $\widehat{\bgamma}_{u, k}$.

The high-dimensional debiased calibration (HDC) estimator of the score ${\E}_{\mathbb{Q}} \{\bS(Y, \bZ; \bbeta)\}$ is constructed as 
\bea
\widehat{S}_{ \HDC,u,j}(\bbeta) &=& \sum_{i \in \mathcal{S}_{\mathbb{F}_u}} \widetilde{p}_{u,i} {S}_{aj}(Y_i, \bZ_i; \bbeta) + \frac{1}{n_0} \sum_{i \in S_{\mathbb{Q}}} {S}_{bj}(\bZ_i, \bbeta), 
\label{eq:AEL-g}
\eea
and the debiased estimate $\widehat{\bbeta}_{\HDC,u}$ of $\bbeta$ using the $u$th source is the solution to
\be
%\widehat{\bbeta}_{\AEL} = \underset{\bbeta}{\argmin} \ 
\widehat{\bS}_{\HDC,u}(\bbeta) = (\widehat{S}_{\HDC, u, 1}(\bbeta), \ldots, \widehat{S}_{\HDC, u, q}(\bbeta))^{\T} = \bzero. 
\label{eq:obj-HD}
\ee
%We propose to solve (\ref{eq:obj-HD}) in a proper neighborhood of $\widehat{\bbeta}_{\HC, u}$, taking advantage of the consistency of the initial estimator.
Note that solving (\ref{eq:obj-HD}) is computationally easy, as the HDC score $\widehat{\bS}_{\HDC,u}(\bbeta)$ depends on $\bbeta$ only through ${S}_{aj}(Y_i, \bZ_i; \bbeta)$ and ${S}_{bj}(\bZ_i, \bbeta)$. The calibration weights $\{\widetilde{p}_{u,i}\}$ are free of $\bbeta$. 
% As $\widetilde{S}_j(Y_i, \bZ_i; \bbeta)$, $\widetilde{S}_{j2}(\bZ_i, \bbeta)$ and the regression coefficients $\widetilde{\btheta}_{u,j,h}$ all depend on $\bbeta$, an iterative algorithm can be used to solve (\ref{eq:obj-HD}). 
% Given the estimate $\widehat{\bbeta}_{\AEL, u, t}$ from the $t$th step, obtain the regression coefficients $\widetilde{\btheta}_{u,j,h, t} = \widetilde{\btheta}_{u,j,h}(\widehat{\bbeta}_{\AEL, u, t})$ from (\ref{eq:EL-outcome}) with $\bbeta = \widehat{\bbeta}_{\AEL, u, t}$. Then, calculate the updated estimate $\widehat{\bbeta}_{\AEL, u, t+1}$ from solving (\ref{eq:obj-HD}) with the regression coefficients $\widetilde{\btheta}_{u,j,h, t}$ from the previous step, and iterate until converge.

Given estimated DR $\widehat{w}_{u}(\bX_i)$ and OR $\widehat{\E}\{\bS_a(Y,\bZ;\bbeta) \mid \bX\}$ from two working models, the AIPW estimate of the score function takes the form $\widehat{\bS}_{u}(\bbeta) = \sum_{i \in \mathcal{S}_{\mathbb{F}_u}} \widehat{w}_{u}(\bX_i) \big[ {\bS}_{a}(Y_i, \bZ_i; \bbeta) - \widehat{\E}\{\bS_a(Y_i,\bZ_i;\bbeta) \mid \bX_i\} \big] + n_0^{-1} \sum_{i \in S_{\mathbb{Q}}} \big[\widehat{\E}\{\bS_a(Y_i,\bZ_i;\bbeta) \mid \bX_i\} + {\bS}_{b}(\bZ_i, \bbeta) \big]$, where the unknown parameter $\bbeta$ is involved in both ${\bS}_{a}(Y_i, \bZ_i; \bbeta)$ and its estimated conditional mean $\widehat{\E}\{\bS_a(Y,\bZ;\bbeta) \mid \bX\}$. This may bring nonconvex issue in solving $\widehat{\bS}_{u}(\bbeta) = \bzero$, where the derivative of $\widehat{\bS}_{u}(\bbeta)$ is no longer positive definite even if its population value is positive definite. Compared to AIPW estimators, the proposed method would not suffer this issue due to the simple IPW form of the HDC estimator in (\ref{eq:AEL-g}) with positive weights $\{\widetilde{p}_{u,i}\}$ free of $\bbeta$.

Furthermore, the proposed method is capable of incorporating multiple DR models. Multiple OR models $\{m_k(\bX_i, \bxi_{u, k})\}$ for ${\E}_{\mathbb{Q}} \{\bS_{a}(Y_i, \bZ_i; \bbeta) | \bX\}$ can be incorporated as well by balancing $\bS_{a}(m_{k}(\bX_i, \widehat{\bxi}_{u, k}), \bZ_i)$ in (\ref{eq:EL-CB-exact}). 
%The proposed approach can therefore achieve multiple robustness in the sense of consistent estimation if any of those models is correctly specified, as shown in Theorem \ref{thm: multiply robust consistency of Z estimator}. 
However, the AIPW estimate only uses a single model for $w_u(\bx)$ and $\E_{\mathbb{Q}}\{\bS(Y, \bZ; \bbeta) | \bX\}$ respectively, and it is not multiply robust. 
%$\sum_{i \in \mathcal{S}_{\mathbb{F}_u}} p_i \bS_{a}(m_{k}(\bX_i, \widehat{\bxi}_{u, k}), \bZ_i) = n_0^{-1} \sum_{i \in \mathcal{S}_{\mathbb{Q}}} \bS_{a}(m_{k}(\bX_i, \widehat{\bxi}_{u, k}), \bZ_i)$ 

We will show the proposed estimator $\widehat{\bbeta}_{\HDC,u}$ is multiply robust in inference. It is asymptotically normal if one of the working density ratio models $\{w_{u, k}(\bX, \bgamma_{u, k})\}$ is correctly specified for the $u$th source, or the outcome regression model of ${\E}_{\mathbb{Q}} \{\bS_{a}(Y, \bZ; \bbeta_0) | \bX\}$ in (\ref{eq:HD-OR}) is correctly specified 
%for some functions $\{\widetilde{S}_{aj}(Y, \bZ)\}_{j = 1}^{q}$ and $\bb(\bx)$ 
at the true parameter value $\bbeta_0$. 
Those advantages of the HDC estimator are due to carefully designed calibration constraints for an augmented set of covariates: the internal bias correction (IBC) constraint in (\ref{eq:EL-IBC}) for incorporating multiple estimated density ratios, the pre-trained model calibration constraint in (\ref{eq:EL-CB-exact}) for incorporating a pre-trained model of $Y$, the projection constraint in (\ref{eq:EL-CB-projection}) for incorporating the OR model of the score function, and the soft orthogonality constraints in (\ref{eq:EL-CB-HD-singlesource}) for removing the first-order influence of estimating the nuisance parameters $\bxi_u$ and $\bgamma_{u, k}$. Those constraints together
%Conducting the augmented outcome regression and adding the derivatives of $\widetilde{\bS}_{a}(m(\bx, \widehat{\bxi}_u), \bZ)$ and $w_{u, k}^{-1}(\bx,\widehat{\bgamma}_{u, k})$ to the balancing functions $\widehat{\bfd}_u(\bx)$ 
creates the Neyman orthogonal score functions of $\bbeta$ in (\ref{eq:AEL-g}), which leads to the debiased estimate $\widehat{\bbeta}_{\HDC, u}$ based on the initial estimate $\widehat{\bbeta}_{\HC, u}$. The estimation variation of all nuisance parameters has no impact on the asymptotic distribution of $\widehat{\bbeta}_{\HDC, u}$. We have explained this point in detail in Section S3 in the SM.

\setcounter{equation}{0}
\section{Specification Test for Transferability}\label{sec:test}

%As discussed in Section \ref{sec:prelim}, researchers may have access to multiple data sources. 
Given the estimate of $\bbeta$ using data from each source, we want to test the transferability assumption for all sources. Namely, we consider testing the null hypotheses: Condition \ref{as:ignorability} holds for all $\mathbb{F}_{u}$.
%\be
%H_0^{\ast}: \mbox{Condition \ref{as:ignorability} holds for all $\mathbb{F}_{u}$ \ vs. \ } H_a^{\ast}: \mbox{Condition \ref{as:ignorability} doesn't hold for some $\mathbb{F}_{u}$}.
% H_0: \mbox{ $\bbeta_{\AEL,u}^{\ast}$ are the same for all $u \in [g]$ \ vs. \ } H_a: \mbox{ $\bbeta_{\AEL,u}^{\ast}$ is different for some $u$ }.
%\label{eq: transferability Hypotheses}
%\ee
However, this hypothesis is not testable as the response $Y$ is missing in the target population. 
Let $\bbeta_{\AEL,u}^{\ast}$ denote the probability limit of $\widehat{\bbeta}_{\AEL,u}$ if it exists. 
Note that, if Condition \ref{as:ignorability} holds for all source distributions, $\bbeta_{\AEL,u}^{\ast}$ should be the same for all $u$. Therefore, we resort to testing the hypothesis
\be
H_0: \mbox{ $\bbeta_{\AEL,u}^{\ast}$ are the same for all $u \in [g]$ \ vs. \ } H_a: \mbox{ $\bbeta_{\AEL,u}^{\ast}$ is different for some $u$ }.
\label{eq:Hypotheses}
\ee
If $H_0$ is rejected, we have evidence against the transferability assumption. It is important to note that the null hypothesis in (\ref{eq:Hypotheses}) only describes the homogeneity across different sources, 
%in other words, the asymptotic equality of the estimated parameters using each source, 
which is necessary but not sufficient for the transferability hypothesis.

Let $\widehat{\bbeta}_{\AEL} = (\widehat{\bbeta}_{\AEL,1}^{\T}, \ldots, \widehat{\bbeta}_{\AEL,g}^{\T})^{\T}$ be the vector of the proposed single-source HDC estimates from all sources. Let $w_{u,i}^{\ast}$, $\bfd_u^{\ast}(\bx)$, $\ba_{u,1i}^{\ast}$, $\ba_{u,2i}^{\ast}$, and $\btheta_{u, j, h}^{\ast}$ be the probability limits of $n_{1u}\widetilde{p}_{u,i}$, $\widehat{\bfd}_u(\bx)$, $\widehat{\ba}_{u,1i}$, $\widehat{\ba}_{u,2i}$, and $\widetilde{\btheta}_{u,j,h}$ for $u \in [g]$, $j \in [q]$ and $h = 1, 2, 3$, evaluated at $\bbeta_0$. Their rigorous definitions are given in Section S2 in the SM.
% Each $\widehat{\bbeta}_{\AEL,u}$ has a influence function expansion. Show that $\widehat{\bbeta}_{\AEL}$ is jointly asymptotic normal with a non-generate covariance matrix. We can test the mean of $\widehat{\bbeta}_{\AEL}$ being the same by using a contrast matrix $\bC \widehat{\bbeta}_{\AEL}$.
Let $w_u^{\ast}(\bX)$ be the probability limit of $n_{1u}\widetilde{p}_{u, i}$ for the observation with covariate $\bX$. Let 
% $\bV_0 = {\E}_{\mathbb{Q}} \big\{ \frac{\partial}{\partial \bbeta} \bS(Y, \bZ; \bbeta) \big|_{\bbeta = \bbeta_0} \big\}$
\be
\bV_{0,u} = \E_{\mathbb{F}_u} \bigg\{w_u^{\ast}(\bX)\frac{\partial}{\partial\bbeta} {\bS}_a(Y,\bZ;\bbeta) \bigg\vert_{\bbeta = \bbeta_0}  \bigg\} + \E_{\mathbb{Q}} \bigg\{ \frac{\partial}{\partial \bbeta}{\bS}_b(\bZ,\bbeta)  \bigg\vert_{\bbeta = \bbeta_0}  \bigg\}, \label{eq:V0}
\ee
$\psi_{u, 1, j}(Y_i,\bX_i) = w_{u,i}^{\ast} \{{S}_j(Y_i, \bZ_i; \bbeta_0) - \psi_{u, 2, j}(\bX_i)\}$
and 
$\psi_{u, 2, j}(\bX_i) = {S}_{bj}(\bZ_i, \bbeta_0) + {\btheta}_{u,j,1}^{\ast \T} {\bfd}_u^{\ast}(\bX_i) + {\btheta}_{u,j,2}^{\ast \T} {\ba}_{u,1i}^{\ast} + {\btheta}_{u,j,3}^{\ast \T} {\ba}_{u,2i}^{\ast}$, which constitutes the influence function of the HDC estimator $\widehat{\bbeta}_{\AEL, u}$. Let $\psi_{u, 1}(Y,\bX) = (\psi_{u, 1, 1}(Y,\bX), \ldots, \psi_{u, 1, q}(Y,\bX))^{\T}$ and $\psi_{u, 2}(\bX) = (\psi_{u, 2, 1}(\bX), \ldots, \psi_{u, 2, q}(\bX))^{\T}$.
From Theorem \ref{thm: IFs of target parameter}, if all sources satisfy Condition \ref{as:ignorability}, $\widehat{\bbeta}_{\AEL}$ is jointly asymptotic normal as 
\be
\sqrt{n_0}(\widehat{\bbeta}_{\AEL} - \mathbf{1}_g \otimes \bbeta_0) \overset{d}{\to} \mathcal{N}(\mathbf{0}_{gq}, \bSigma_{\AEL} ), \nn 
\ee
where $\otimes$ stands for the Kronecker product, $\mathbf{1}_g$ is a $g$-dimensional vector of 1, $\bSigma_{\AEL} = (\bSigma_{\AEL}^{(u_1u_2)})$ is a $gq\times gq$ block covariance matrix, and $\{ \bSigma_{\AEL}^{(u_1u_2)} \}$ are $q\times q$ matrices of the form
\bea
\bSigma_{\AEL}^{(u u)} &=& \bV_{0,u}^{-1} \big[ c_u^{-1}{\E}_{\mathbb{F}_u}\{\psi_{u, 1}(Y,\bX)\psi_{u, 1}^{\T}(Y,\bX)\} + {\E}_{\mathbb{Q}}\{\psi_{u, 2}(\bX)\psi_{u, 2}^{\T}(\bX)\} \big] \bV_{0,u}^{-\T}, \label{eq:cov-1} \\
\bSigma_{\AEL}^{(u_1 u_2)} &=& \bV_{0,u_1}^{-1} \big[ {\E}_{\mathbb{Q}}\{\psi_{u_1, 2}(\bX) \psi_{u_2, 2}^{\T}(\bX)\} \big] \bV_{0,u_2}^{-\T} \label{eq:cov-2}
\eea
for $u, u_1, u_2 = 1, \ldots, g$, and $c_u = \lim n_{1u} / n_0$.

% with $u$-th diagonal being $ \frac{W_u}{W}V_{\bbeta_0}^{-1}\E\psi_u(Y,\bX,D;\bbeta_0)\psi_u(Y,\bX,D;\bbeta_0) V_{\bbeta_0}^{-1}$, and off-diagonals,corresponding to the common supervised samples in the asymptotic expansion, being $\frac{1}{W} V_{\bbeta_0}^{-1}\V_{\mathbb{Q}}\{\psi_{(2)}(Y,\bX;\bbeta_0)\}V_{\bbeta_0}^{-1}$, 
%\be
%\Sigma_{\AEL} = \frac{1}{W}
%\begin{pmatrix}
%    W_1\Sigma_{\AEL,1} & \Sigma_{\AEL,\text{off}} & \Sigma_{\AEL,\text{off}} & \cdots & \Sigma_{\AEL,\text{off}} \\
%    \Sigma_{\AEL,\text{off}} & W_2\Sigma_{\AEL,2} & \Sigma_{\AEL,\text{off}} & \cdots & \Sigma_{\AEL,\text{off}} \\
%    \vdots & \vdots & \vdots & \ddots & \vdots \\
%    \Sigma_{\AEL,\text{off}} & \cdots & \cdots & \cdots & W_g\Sigma_{\AEL,g}
%\end{pmatrix} \nn
%\ee

For the hypotheses in (\ref{eq:Hypotheses}), we test the  asymptotic means of $\widehat{\bbeta}_{\AEL, u}$ being the same for all $u = 1, \ldots, g$.
Let $\bC = (\bone_{g-1},-\bI_{g-1})\otimes \bI_q$
%Then we can test the mean of $\widehat{\bbeta}_{\AEL}$ being the same by using a contrast matrix $\bC \widehat{\bbeta}_{\AEL}$
%\be
%\bC = 
%\begin{pmatrix}
%I_q & - I_q & \bzero & \cdots &\bzero \\
%I_q & \bzero & - I_q &  \cdots & \bzero \\ 
%\vdots & \vdots & \vdots &\ddots &\vdots \\ 
%I_q & \bzero &\cdots &\bzero & -I_q
%\end{pmatrix}\in \mathbb{R}^{(g-1)q \times gq} \nn
%\ee
%\be
%\bC = 
% \begin{pmatrix}
% \bI_q & -\bI_q & \cdots &\bzero \\
% \vdots & \vdots &\ddots &\vdots \\ 
% \bI_q & \bzero & \cdots & -\bI_q
% \end{pmatrix}
%\in \mathbb{R}^{(g-1)q \times gq} \nn
%\ee
be a contrast matrix, where $\bI_q$ denotes the $q$-dimensional identity matrix. 
Under the null hypothesis, Proposition \ref{prop: specification test} shows that the test statistic 
\be
F = n_0 (\bC\widehat{\bbeta}_{\AEL})^{\T}(\bC\widehat{\bSigma}_{\AEL}\bC^{\T}  )^{-1}(\bC\widehat{\bbeta}_{\AEL}) \nn
\ee
is asymptotically $\chi^2_{(g-1)q}$ distributed with $(g-1)q$ degrees of freedom, where $\widehat{\bSigma}_{\AEL}$ is the plug-in estimate of $\bSigma_{\AEL}$, constructed in (\ref{eq: confidence interval}). We reject the null hypothesis of (\ref{eq:Hypotheses}) if $F > \chi^2_{(g-1)q}(\alpha)$, where $\chi^2_{(g-1)q}(\alpha)$ is the upper $\alpha$ quantile of the distribution $\chi^2_{(g-1)q}$.

\setcounter{equation}{0}
\section{Multi-source Inference}\label{sec:inference}

If there is no evidence of rejecting the hypotheses in (\ref{eq:Hypotheses}), we would like to combine the information from multiple sources to obtain an efficient estimator of $\bbeta$ on the target population. 
One possible way is to apply the estimating equation (\ref{eq:obj-HD}) to each source with different high-dimensional working density ratio models $w_{u,k}(\bX,\bgamma_{u,k})$ and regression model $m(\bX,\bxi_u)$ for $u \in [g], k\in [v_u]$ to obtain estimators $\widehat{\bbeta}_{\AEL,1}, \ldots, \widehat{\bbeta}_{\AEL,g}$. A naive estimator of $\bbeta$ that combines multi-source information is the weighted estimator $\widehat{\bbeta}_{\A} = \sum_{u=1}^g \bA_u \widehat{\bbeta}_{\AEL, u}$,
where $\{\bA_u\}_{u = 1}^{g}$ are $q \times q$ weighting matrices satisfying $\sum_{u=1}^g \bA_u = \bI_q$. Let $\bA = (\bA_1, \ldots, \bA_g) \in \mathbb{R}^{q\times qg}$. Note that $n_0 \var(\widehat{\bbeta}_{\A}) = \bA \bSigma_{\AEL} \bA^{\T} \{1 + o(1)\}$. To minimize the main order of the variances of $\widehat{\bbeta}_{\A}$, we choose the weights 
\be\label{eq:opt-weight}
\bA^{\ast} = \argmin \ \tr(\bA \bSigma_{\AEL} \bA^{\T}) \mbox{ \ subject to \ } \sum_{u=1}^g \bA_u = \bI_q.
\ee
Let $\bJ = (\bI_q, \ldots, \bI_q)^{\T} \in \mathbb{R}^{qg \times q}$. 
%\citet{Lavancier2016Combineestmators} 
It can be showed that the optimal weights are $\bA^{\ast} = (\bJ^{\T} \bSigma_{\AEL}^{-1} \bJ)^{-1} \bJ^{\T} \bSigma_{\AEL}^{-1}$, which leads to the optimal weighted estimator
\be\label{eq:combined-est}
\widehat{\bbeta}_{\A, \opt} = \sum_{u=1}^g \bA_u^{\ast} \widehat{\bbeta}_{\AEL, u},
\ee
and its covariance matrix $n_0 \var(\widehat{\bbeta}_{\A,\opt}) = (\bJ^{\T} \bSigma_{\AEL}^{-1} \bJ)^{-1}\{1 + o(1)\}$.

We now propose a new method to combine information across multiple sources and obtain a more efficient estimator than $\widehat{\bbeta}_{\A, \opt}$ and all naive weighted estimators. Rather than balancing the augmented covariate functions between the target group $\mathcal{S}_{\mathbb{Q}}$ and each source group $\mathcal{S}_{\mathbb{F}_u}$ separately as in (\ref{eq:EL-CB-HD-singlesource}), we balance those covariate functions, including all DR models and OR working models, between $\mathcal{S}_{\mathbb{Q}}$ and the mixture group 
$\mathcal{S}_{\mathbb{F}} = \bigcup_{u=1}^g \mathcal{S}_{\mathbb{F}, u}$
of all sources. 
%By leveraging shared covariates, the covariate balancing technique can be systematically applied to both density ratio and regression models across the entire mixed source dataset, rather than being constrained to individual specific sources. 
It facilitates information sharing across sources, utilizing density ratios from all sources for bias calibration and improving the efficiency of estimating $\bbeta_0$. 
%For example, SCB (\ref{eq:EL-CB-HD-singlesource}) for source-specific basis functions can be augmented to the union of all training data sources
%\be
%\bigg| \sum_{i \in \mathcal{S}_{\mathbb{F}_u}} p_{u,i} \widehat{\bfd}_{u,c}(\bX_i) \bigg|_{\infty} \leq \omega_{\ps} \ 
%\Rightarrow \   \bigg|\sum_{i \in  \mathcal{S}_{\mathbb{F}}} p_i \widehat{\bfd}_{u,c}(\bX_i) \bigg|_{\infty} \leq \omega_{\ps} : u \in [g] \nn 
%\ee

Specifically, we combine the functions $\widehat{\bfd}_u(\bx)$ for each source in (\ref{eq:basis-HD}) to an augmented set of functions for all sources:
\be
\widehat{\bfd}(\bx) = \big(\bb^{\T}(\bx), \widetilde{\bS}_{a}^{(1) \T}(\bx; \widehat{\bxi}_u), {w_{u, k}^{(1)\T}(\bx,\widehat{\bgamma}_{u, k})}{ w_{u, k}^{-2}(\bx,\widehat{\bgamma}_{u, k})}  : u \in [g], k \in [v_u] \big)^{\T}.
\label{eq:basis-HD-multi-source}
\ee 
% the original basis functions and the first order derivatives of the predicted score $\bS_{u,j}^{(1)}(\bX; \widehat{\bxi}_u, \bbeta)$ and the inverse PS $\{1 - \pi_{u, k}(\bX, \widehat{\bgamma}_{u, k})\}^{-1}$ for all possible $u$, $j$ and $k$. 
%where the polynomials $\{\pi_{k}^{l}(\bX, \widehat{\bgamma}_k): 0 \leq l \leq u \}$ are used to approximate the functional and derivatives of $\pi_{k}(\bX, \bgamma_k)$. Here, $u$ can be chosen at the order $\log(n)$ as $\pi_{k}^{l}(\bX, \widehat{\bgamma}_k)$ decays exponentially fast to zero with the increase of $l$. Note that restricting $l = 0$ corresponds to the original basis function $\bb(\bX)$. 
Let $\widehat{\bw}_i = (\widehat{\bw}_{1,i}^{\T},\ldots,\widehat{\bw}_{g,i}^{\T})^{\T} =(w_{1,1}(\bX_i, \widehat{\bgamma}_{1,1}), \ldots, w_{g, v_g}(\bX_i, \widehat{\bgamma}_{g,v_g}))^{\T}$, $\bar{\bfd}_0 = n_0^{-1} \sum_{i \in \mathcal{S}_{\mathbb{Q}}} \widehat{\bfd}(\bX_i)$ and $\widehat{\bfd}_{c}(\bX_i) = (\widehat{d}_{c, 1}(\bX_i), \ldots, \widehat{d}_{c, \widetilde{m}}(\bX_i))^{\T} = \widehat{\bfd}(\bX_i) - \bar{\bfd}_0$, where $\widetilde{m}$ is the dimension of $\widehat{\bfd}(\bX)$. 
%Let $|\cdot|_{\infty}$ denote the element-wise maximum norm of a vector or matrix.
%For the high-dimensional scenario where the number of covariates is large, exact covariate balancing in (\ref{eq:EL-CB-fixedD}) is not feasible. We need to develop a new method for high-dimensional semi-supervised learning. Recall that $\widehat{\bgamma}_k$ is the estimate of $\bgamma_k$ and $\pi_{k}^{(1)}(\bX, \bgamma_k)$ is the first order derivative of the $k$th working model $\pi_k(\bX, \bgamma_k)$. 
%Given an initial estimate $\widetilde{\bbeta}$, 

Similar to (\ref{eq:EL-outcome}), given an initial estimate $\widehat{\bbeta}^{\ini}$, let $(\widetilde{\btheta}_{j,1}, \widetilde{\btheta}_{j,2}, \widetilde{\btheta}_{j,3})$ be the estimated coefficients of the augmented regression of $S_{aj}(Y_i, \bZ_i; \widehat{\bbeta}^{\ini})$ over $\mathcal{S}_{\mathbb{F}}$ with weights $\{\widehat{p}_i\}_{i\in S_{\mathbb{F}}}$:
\bea
(\widetilde{\btheta}_{j,1}^{\T}, \widetilde{\btheta}_{j,2}^{\T}, \widetilde{\btheta}_{j,3}^{\T}) \ =
&\underset{\btheta_{j,1}, \btheta_{j,2}, \btheta_{j,3}}{\operatorname{argmin}} &\frac{1}{2n_1} \sum_{i \in \mathcal{S}_{\mathbb{F}}} (n_1\widehat{p}_i)^{2} \{ {S}_{aj}(Y_i, \bZ_i; \widehat{\bbeta}^{\ini}) - \btheta_{j,1}^{\T} \widehat{\bfd}(\bX_i) \nn \\ 
&&  \ 
- \btheta_{j,2}^{\T} \widehat{\ba}_{1i} - \btheta_{j,3}^{\T} \widehat{\ba}_{2i} \}^2  + \omega_{\oc} 
(|\btheta_{j,1}|_1 + |\btheta_{j,2}|_1 + |\btheta_{j,3}|_1) \ \ \  \label{eq:EL-outcome-multi-source}
\eea
for $j = 1, \ldots, q$, where $\{\widehat{p}_i\}_{i\in S_{\mathbb{F}}}$ is the solution to the multi-source empirical likelihood problem in (\ref{eq:EL-observe-multi}) with solely the IBC constraints in (\ref{eq:EL-IBC-multi-source}), $\widehat{\ba}_{1i} = (\widehat{\ba}_{1,1i}^{\T}, \ldots, \widehat{\ba}_{g,1i}^{\T})^{\T}$ with $\widehat{\ba}_{u,1i} = \widetilde{\bS}_a(m(\bX_i, \widehat{\bxi}_u), \bZ_i)$, $\widehat{\ba}_{2i} =\widehat{\bw}_i^{-1}$ and $\omega_{\oc}$ is a penalty parameter for the outcome regression. The regressor functions $\widehat{\ba}_{1i}$ and $\widehat{\ba}_{2i}$ are augmented to include their components from each source. Here, the initial estimate can be chosen as the single-source estimate $\widehat{\bbeta}_{\AEL, u}$ from any of the sources.

Let $\widehat{\bU}_{i} =(\widehat{\bfd}(\bX_i)^{\T},\widehat{\ba}_{1i}^{\T},\widehat{\ba}_{2i}^{\T})^{\T}$ and $\widetilde{\btheta}_{j} = (\widetilde{\btheta}_{j,1}^{\T}, \widetilde{\btheta}_{j,2}^{\T}, \widetilde{\btheta}_{j,3}^{\T})^{\T}$. The calibration weights $\widetilde{p}_i$ over the mixture group $\mathcal{S}_{\mathbb{F}}$ are calculated from the constrained optimization:
\bea
\{\widetilde{p}_i: i \in \mathcal{S}_{\mathbb{F}}\} &=& \underset{ \{p_i: i \in \mathcal{S}_{\mathbb{F}}\} }{\operatorname{argmax}} \sum_{i \in \mathcal{S}_{\mathbb{F}}} \log(p_i), \mbox{ \ subject to \ } p_i \geq 0, \ \sum_{i \in \mathcal{S}_{\mathbb{F}}} p_i = 1, \label{eq:EL-observe-multi} \\
&&  \sum_{i \in \mathcal{S}_{\mathbb{F}}} p_i w_{u, k}^{-1}(\bX_i, \widehat{\bgamma}_{u, k})   = 1 \mbox{ \ for $u \in [g], k \in [v_u]$,} \ \ \ \ \ \ \  \label{eq:EL-IBC-multi-source} \\ 
&& \sum_{i\in S_{\mathbb{F}}} p_i \widehat{\bU}_{i}^{\T}\widetilde{\btheta}_j = \frac{1}{n_0}\sum_{i\in S_{\mathbb{Q}}} \widehat{\bU}_{i}^{\T}\widetilde{\btheta}_j  \mbox{ \ for $j = 1, \ldots, q$,} 
\label{eq:EL-CB-projection-multi} \\
&& \sum_{i \in \mathcal{S}_{\mathbb{F}}} p_i \{ \widetilde{\bS}_a(m(\bX_i, \widehat{\bxi}_u), \bZ_i) - \bar{\bS}_{a,u} \} = 0 \mbox{ \ for $u \in [g]$}, \label{eq:EL-PT-HD-multi-source} \\
&& \bigg| \sum_{i \in \mathcal{S}_{\mathbb{F}}} p_i \widehat{\bfd}_c(\bX_i) \bigg|_{\infty} \leq \omega_{\ps}. \label{eq:EL-CB-HD-multi-source}
\eea
Similar to the single source problem in (\ref{eq:EL-observe-inference}), (\ref{eq:EL-IBC-multi-source}) is the IBC constraints that calibrate the weights to the density ratio $w(\bx)$ of $\mathbb{Q}_{\bX}$ over the mixture distribution $\mathbb{F}_{\bX}$ using the DR models of all sources, (\ref{eq:EL-CB-projection-multi}) and (\ref{eq:EL-PT-HD-multi-source}) are the multi-source version of the projection and pre-trained model calibration constraints respectively, and (\ref{eq:EL-CB-HD-multi-source}) is the soft orthogonality constraints for the HD augmented covariates in (\ref{eq:basis-HD-multi-source}) over the mixture group $\mathcal{S}_{\mathbb{F}}$. 
% where $|\cdot|_{\infty}$ denotes the element-wise maximum norm of a vector or matrix, and $\bar{\bS}_{u, 0} = n_0^{-1} \sum_{i \in \mathcal{S}_{\mathbb{Q}}} \bS(m_u(\bX_i, \widehat{\bxi}_u), \bZ_i; \bbeta)$. 
%It is well known that the IPW estimator with the covariate balancing property is equivalent to the AIPW estimator. However, for the high-dimensional balancing functions $\widehat{\bfd}(\bX)$ with the soft calibration constraints in (\ref{eq:EL-CB-HD}), $\sum_{i = 1}^{n} D_i \widetilde{p}_i g(Y_i, \bZ_i; \bbeta)$ is no longer an AIPW estimator due to the non-exact calibration. 
% To correct the bias from the soft calibration in (\ref{eq:EL-CB-HD}), we consider estimating the outcome regression of ${\E}_{\mathbb{Q}}\{\bS(Y, \bZ; \bbeta) \vert \bX\}$ by minimizing the regularized weighted augmented linear regression with the weight $(n_1\widetilde{p}_i)^{2}$ on the source samples\footnote{only L1 penalty?}:

% It is shown in the SM that $(n_1\widetilde{p}_i)^{2} \to w^2(\bX_i)$ as $n, p \to \infty$ if one of the working PS models $\{\pi_{u,k}(\bX, \bgamma_{u,k})\}$ is correctly specified for each source.
	
The multi-source high-dimensional debiased calibration (MHDC) estimators of the score functions ${\E}_{\mathbb{Q}} \{\bS(Y, \bZ; \bbeta)\}$ can be constructed as 
% \bea
% \widehat{S}_{\MDEL,j}(\bbeta) &=& \sum_{i \in \mathcal{S}_{\mathbb{F}}} \widetilde{p}_i \{{S}_{aj}(Y_i, \bZ_i; \bbeta) - \widetilde{\btheta}_{j,1}^{\T} \widehat{\bfd}(\bX_i) - \widetilde{\btheta}_{j,2}^{\T} \widehat{\ba}_{1i} - \widetilde{\btheta}_{j,3}^{\T} \widehat{\ba}_{2i}\} \nn \\
% &+& \frac{1}{n_0} \sum_{i \in \mathcal{S}_{\mathbb{Q}}} \{ \widetilde{\btheta}_{j,1}^{\T} \widehat{\bfd}(\bX_i) + \widetilde{\btheta}_{j,2}^{\T} \widehat{\ba}_{1i} + \widetilde{\btheta}_{j,3}^{\T} \widehat{\ba}_{2i} \} + \frac{1}{n_0}\sum_{i\in S_{\mathbb{Q}}} {S}_{bj}(\bZ_i,\bbeta),
% \label{eq:AEL-g-multi}\eea
\bea
\widehat{S}_{\MHDC,j}(\bbeta) &=& \sum_{i \in \mathcal{S}_{\mathbb{F}}} \widetilde{p}_i \{{S}_{aj}(Y_i, \bZ_i; \bbeta) \}+ \frac{1}{n_0}\sum_{i\in S_{\mathbb{Q}}} {S}_{bj}(\bZ_i,\bbeta),
\label{eq:AEL-g-multi}\eea
and the MHDC estimate $\widehat{\bbeta}_{\MHDC}$ of $\bbeta$ is the solution to
\be
%\widehat{\bbeta}_{\AEL} = \underset{\bbeta}{\argmin} \ 
\widehat{\bS}_{\MHDC}(\bbeta) = (\widehat{S}_{\MHDC,1}(\bbeta), \ldots, \widehat{S}_{\MHDC,q}(\bbeta))^{\T} = \bzero.
\label{eq:obj-HD-multi}\ee

% Similar to (\ref{eq:obj-HD}), solving the MHDC scores being zero in (\ref{eq:obj-HD-multi}) is computationally easy as all the terms $\{\widetilde{\btheta}_{j,1}^{\T} \widehat{\bfd}(\bX_i) + \widetilde{\btheta}_{j,2}^{\T} \widehat{\ba}_{1i} + \widetilde{\btheta}_{j,3}^{\T} \widehat{\ba}_{2i}\}$ are free of $\bbeta$.

In the next section, we will show that the proposed MHDC estimator $\widehat{\bbeta}_{\MHDC}$ provides multiply robust inference for the target parameter $\bbeta_0$, and it is more efficient than the optimally weighted estimator $\widehat{\bbeta}_{\A, \opt}$ in (\ref{eq:combined-est}).
Similar to the discussion at the end of Section \ref{sec: single source inference}, the proposed MHDC procedure achieves Neyman orthogonality of nuisance parameters from all sources to the target parameter by carefully designing the augmented set of calibration functions in (\ref{eq:basis-HD-multi-source}) and the augmented outcome regression in (\ref{eq:EL-outcome-multi-source}).
%such that the estimation of the nuisance parameters from all sources would not affect the asymptotical distribution of $\widehat{\bbeta}_{\AEL}$. See the detailed explanations in Section \ref{SM:intuition} in the SM.

%Let $\{\bgamma_{k, \ast}\}_{k = 1}^{q}$, $\lambda_{0, \ast}$, $\blambda_{1, \ast}$, $\blambda_{2, \ast}$, $\bbeta_{1, \ast}$, $\bbeta_{2, \ast}$ be the probability limits of $\{\widehat{\bgamma}_k\}_{k = 1}^{q}$, $\widetilde{\lambda}_{0}$, $\widetilde{\blambda}_{1}$, $\widetilde{\blambda}_{2}$, $\widetilde{\bbeta}_{1}$, $\widetilde{\bbeta}_{2}$ as $n, p \to \infty$, respectively.

\setcounter{equation}{0}
\section{Asymptotic Results}\label{sec:asym}

% theta^{ast}  depend on beta 与sec5不同 ?
% 是否能用beta^{ini} 有待考证
% theta^{\ast}(\beta) condition 把beta加上
% multisource 6.7后面 theta depend on beta but for simplicity we drop it when there is 

In this section, we state the asymptotic theories for the proposed estimators. 
%While the conditions are initially stated for a single source, we extend the result to the multi-source scenario by assuming that they hold for all sources.
% First, we restate the probability limits of the estimated nuisance parameters.
Let $\bgamma_{u,k}^{\ast}$, $\bxi_u^{\ast}$, $w_{u,i}^{\ast}$, $w_{i}^{\ast}$, $\btheta_{u,j,h}^{\ast}$ and $\btheta_{j,h}^{\ast}$ be the probability limits of $\widehat{\bgamma}_{u,k}$, $\widehat{\bxi}_u$, $n_{1u}\widetilde{p}_{u,i}$, $n_{1}\widetilde{p}_{i}$, $\widehat{\btheta}_{u,j,h}$ and $\widehat{\btheta}_{j,h}$ when $n, p \to \infty$. 
Let $\blambda^{\ast}_{u,1}$, $\blambda^{\ast}_{u,2}$, $\blambda^{\ast}_{u,3}$ and $\blambda^{\ast}_{u,4}$ denote the probability limits of the estimated Lagrange parameters, $\widehat{\blambda}_{u,1}$, $\widehat{\blambda}_{u,2}$, $\widehat{\blambda}_{u,3}$ and $\widehat{\blambda}_{u,4}$ in the dual problem, corresponding to the IBC, SCB, PMC and PC constraints. 
% To simplify notations, we don't distinguish $\blambda^{\ast}_{u,h}$, $h = 1, 2, 3,4$, for the single-source EL problems in (\ref{eq:EL-observe}) and (\ref{eq:EL-observe-inference}). 
The rigorous definitions of those limiting values are detailed in Section S2 of the SM.
% , are derived from population estimating equations.
% \ref{sm: definitions of limiting values} 
The probability limits $\widetilde{\bS}_a(m(\bX,\bxi_u^{\ast}),\bZ )$, $\bfd_u^{\ast}(\bX)$, $\ba_{u,1i}^{\ast}$, $\ba_{u,2i}^{\ast}$,$\bU_{u,i}^{\ast}$, $\bfd^{\ast}(\bX)$, $\ba_{1i}^{\ast}$, $\ba_{2i}^{\ast}$ and $\bU_i^{\ast}$ are obtained by substitution.
% plugging the limiting values of the estimated parameters. 
Denote $a \vee b = \max(a,b)$ and $s_{\lambda u} = |\blambda_{u,2}^{\ast}|_0$ as the sparsity level.
% of the conjugate parameters.
We impose the following regularity conditions.

\begin{as}\label{as: sub gaussian of basis functions and covariates}
Under both the source and target population, $\mathbb{F}_u$ and $\mathbb{Q}$, the covariates $\bX$ and $\bb(\bX)$ are sub-Gaussian random vectors, and $\widetilde{\bS}_{a}(m(\bX,\bxi_u^{\ast}),\bZ)$ is a sub-exponential random vector. 
% There is a positive constants $C$ such that $\forall t \geq 0$, for $\bv \in \mathbb{R}^{m}, \bu \in \mathbb{R}^{p}$ satisfying $|\bu|_2 = |\bv|_2 = 1$ , we have $\mathbb{P}(|\bv^T \bb(\bX)| \geq t) \leq 2 \exp(-t^2/C^2)$ and $\mathbb{P}(|\bu^T \bX| \geq t) \leq 2 \exp(-t^2/C^2)$. 
\end{as}
	
\begin{as}\label{as: bounded eigenvalues}
Let $\bSigma_u = \operatorname{Cov}_{\mathbb{F}_u}\{(\bfd_u^{\ast \T}(\bX),\ba_{u,1}^{\ast \T}(\bX),\ba_{u,2}^{\ast \T}(\bX) )^{\T}\}$
and $\Sigma_{x} = \operatorname{Cov}_{\mathbb{F}_u}(\bX)$. Assume their minimum and maximum eigenvalues are bounded on the source population $\mathbb{F}_u$ such that $1/C \leq \lambda_{\min}(\bSigma_u), \lambda_{\min}(\bSigma_x), \lambda_{\max}(\bSigma_u), \lambda_{\max}(\bSigma_x) \leq C$ for a positive constant $C$.
 % $\bSigma := \operatorname{Cov}\{(\bX,\bfd^{\ast}(\bX),\pi^{\ast})^T\}$ satisfy $C \leq \lambda_{\min}(\bSigma) \leq \lambda_{\max}(\bSigma) \leq 1/C$
\end{as}
% add term X since we need this in theorem 1, debiasing procesure the second derivative of \pi contains XX^T
	
\begin{as}\label{as: sub gaussian of residual of outcome regression}
The residuals $\epsilon_{u,j,i}^{\ast} = {S}_{aj}(Y_i, \bZ_i; \bbeta_0) - {\btheta}_{u,j,1}^{\ast \T} {\bfd}_u^{\ast}(\bX_i) - {\btheta}_{u,j,2}^{\ast \T} {\ba}_{u,1i}^{\ast} - {\btheta}_{u,j,3}^{\ast \T} {\ba}_{u,2i}^{\ast}$ are sub-Gaussian distributed on the source data $\mathbb{F}_u$. 
%There is a positive constant $C_{\epsilon}$ such that  $\forall t \geq 0$,$\PP(|\epsilon_{u,j,i}^{\ast}| \geq t) \leq 2\exp(-t^2/C_{\epsilon}^2)$. 
% When the outcome model is misspecified, the residuals are bounded.
\end{as}
	
\begin{as}\label{as: convergence of DR parameter gamma}
The DR models admit the form $w_{u,k}(\bX,\bgamma_{u,k}) = g_{u,k}(\bX^{\T}\bgamma_{u,k}) $ for $g_{u,k}(\cdot)$ with bounded second-order derivatives such that 
%are bounded in the sense that there exists a positive constant $C$ such that 
$C_1^{-1} \leq w_{u,k}(\bX,\bgamma_{u,k}^{\ast}) \leq C_1 $ 
% $\E\{D_i \pi_k(\bX_i,\bgamma_{k}^{\ast}) \} = \pi(\bX_i) (or \frac{n_1}{n}) , k = 1,\dots, v$ 
% $\E \{ \pi_{u,i}(\bX,\bgamma_{u,i}^{\ast}) \} = \PP_{\mathbb{F}_u,\mathbb{Q}}(D = 1)$
and $\E_{\mathbb{Q}} \{ w_{u,k}^{-1}(\bX,\bgamma_{u,k}^{\ast}) \} $ $
= 1$ for a positive constant $C_1$ and all $k\in [v_u]$.  
The estimators of the parameters satisfy $| \bgamma_{u,k}^{\ast} - \widehat{\bgamma}_{u,k}|_1 \leq C s_{1u}\sqrt{\log(p)/n }  $ and $| \bgamma_{u,k}^{\ast} - \widehat{\bgamma}_{u,k}|_2 \leq C \sqrt{s_{1u}\log(p)/n } $ with probability approaching 1, where  $s_{1u} = \sup_{k \in [v_u]} |\bgamma_{u,k}^{\ast}|_0$.
   %$|\E_n \{\pi_k(\bX_i,\bgamma_{k}^{\ast}) - \pi_k(\bX_i,\hat{\bgamma}_k)   \}| \leq C \sqrt{s_1\log(p)/n}$ for $1 \leq k\leq v$.   Every components of first order derivative of $\pi_k^{(1)}$ is lipschitz continuous and the prediction error satisfies $|\E_n \{\pi_k^{(1)}(\bX_i,\bgamma_k) - \pi_k^{(1)}(\bX_i,\hat{\bgamma}_k)  \}| \leq C \sqrt{s_1\log(p)/n} $. 
If all the density ratio models are misspecified, the limiting values of the calibration weights $n_{1u}\widetilde{p}_{u,i}$ satisfy $C_1^{-1} \leq w_{u,i}^{\ast} \leq C_1$ for all $i$. 
%where $s_{3u}$ is defined in Condition \ref{as: convergence of pretrained model}.
   % the empirical likelihood weights converge to some $\pi_i^{\ast}$ such that $|\E_n (\pi_i^{\ast} - \Tilde{\pi_i})^2| = o_p(1)$ where $\Tilde{\pi_i} = (n_1\tilde{p_i})^{-1}$
\end{as}

\begin{as}\label{as: convergence of pretrained model}
The pre-trained model satisfies $m(\bX,\bxi) = M(\bX^{\T}\bxi)$ for a function $M(\cdot)$ with bounded third-order derivatives.
% \footnote{or lipschitz continuous 2nd order derivative}.
The estimator $\widehat{\bxi}_u$ satisfies $| \widehat{\bxi}_u - \bxi_u^{\ast}|_1 \leq C s_{3u} \sqrt{\log(p)/n}  $ and $| \widehat{\bxi}_u - \bxi_u^{\ast}|_2 \leq C \sqrt{s_{3u}\log(p)/n} $ with probability approaching 1, where $s_{3u} = |\bxi_u^{\ast}|_0$.
\end{as}

\begin{as} \label{as: estimating function}
The estimating function $\bS(Y,\bZ; \bbeta)$ satisfies:
(i) The true parameter $\bbeta_0 \in \text{int}(\mathcal{B})$ for a compact set  $\mathcal{B}$ is the unique solution to $\E_{\mathbb{Q}} \{ \bS(Y,\bZ; \bbeta)  \} = \bzero $.
(ii) $\bS(Y,\bZ; \bbeta)$ has bounded third-order derivatives with respect to $Y$.
(iii) $\bS(Y,\bZ; \bbeta)$ and ${\bS}_b(\bZ,\bbeta)$ are continuously differentiable with respect to $\bbeta$ in a neighborhood of $\bbeta_0$, 
with matrices $\E_{\mathbb{Q}} \{ \frac{\partial}{\partial \bbeta} \bS(Y,\bZ; \bbeta) \vert_{\bbeta = \bbeta_0} \} $, $\E_{\mathbb{F}_u} \big\{w_u^{\ast}(\bX)\frac{\partial}{\partial\bbeta} {\bS}_a(Y,\bZ;\bbeta) \big\vert_{\bbeta = \bbeta_0}  \big\} + \E_{\mathbb{Q}} \big\{ \frac{\partial}{\partial \bbeta}{\bS}_b(\bZ,\bbeta)  \big\vert_{\bbeta = \bbeta_0}  \big\}$
of full rank.
% and  
% $\E_{\mathbb{F}} \{\underset{\bbeta \in \mathcal{B}}{\sup} | \frac{\partial}{\partial\bbeta} \bS(Y,\bZ; \bbeta)|_2 \} < \infty$,
% $\E \{\underset{\bbeta \in \mathcal{B}}{\sup} |  \bS(Y,\bZ; \bbeta)| \} < \infty$
(iv) $\E \{{\sup}_{\bbeta \in \mathcal{B}} | {\bS}_a(Y,\bZ; \bbeta)|_2 \} $ and  
$\E \{{\sup}_{\bbeta \in \mathcal{B}} | {\bS}_b(\bZ, \bbeta)|_2 \}$ are finite.
% $\E \{\underset{\bbeta \in \mathcal{B}}{\sup} |  \bS(Y,\bZ; \bbeta)|_2^2 \} $\footnote{for convergence in (3.6)},\linebreak
% $\E \{\underset{\bbeta \in \mathcal{B}}{\sup} | \frac{\partial}{\partial\bbeta} \bS^{(1)}(\bX,\bxi_u^{\ast};\bbeta)|_2 \} $, 
% $\E \{\underset{\bbeta \in \mathcal{B}}{\sup} | \frac{\partial}{\partial\bbeta} \bS(\bX,\bxi_u^{\ast};\bbeta)|_2 \} $,
% $\E \{\underset{\bbeta \in \mathcal{B}}{\sup} |  \bS(\bX,\bxi_u^{\ast};\bbeta)| \}$, 
(v) For a positive constant $\delta$, $\E \{{\sup}_{|\bbeta-\bbeta_0|\leq \delta} | \frac{\partial}{\partial\bbeta} {\bS}_a(Y,\bZ; \bbeta)|_2 \}$ and 
$\E \{{\sup}_{|\bbeta-\bbeta_0|\leq \delta} | \frac{\partial}{\partial\bbeta} {\bS}_{b}(\bZ, \bbeta)|_2 \}$ are finite . 
% $\E_{\mathbb{F}} \{\underset{\bbeta \in \mathcal{B}}{\sup} | \frac{\partial}{\partial\bbeta} \bS^{(1)}(\bX,\bxi_u^{\ast};\bbeta)|_2 \} < \infty$, 
% $\E_{\mathbb{F}} \{\underset{\bbeta \in \mathcal{B}}{\sup} | \frac{\partial}{\partial\bbeta} \bS(\bX,\bxi_u^{\ast};\bbeta)|_2 \} < \infty$,
% $\E \{\underset{\bbeta \in \mathcal{B}}{\sup} |  \bS(\bX,\bxi_u^{\ast};\bbeta)| \} < \infty$, 
\end{as}

\begin{as}\label{as: sparsity of gamma and theta}
    We assume
    $\sqrt{s_{1u}\vee s_{3u}}(s_{1u}\vee s_{2u} \vee s_{3u})\log(\widetilde{m}_u)\log^{1/2}(p)\log^{1/2}(n)= o(n^{1/2})$ and $(s_{1u}\vee s_{3u})s_{\lambda u}\log(\widetilde{m}_u) = o(n)$, where $s_{2u} = {\sup}_{j \in [q]}|(\btheta_{u,j,1}^{\ast \T},\btheta_{u,j,2}^{\ast\T}, \btheta_{u,j,3}^{\ast\T}) |_0$ denotes the sparsity of the augmented outcome regression.
\end{as}

Conditions \ref{as: sub gaussian of basis functions and covariates}, \ref{as: bounded eigenvalues} and \ref{as: sub gaussian of residual of outcome regression} are standard high-dimensional assumptions ensuring concentration and restricted eigenvalue properties
% conventional regularity conditions for the concentration inequality and the restricted eigenvalue condition in high-dimensional statistics 
\citep{Bickel2008SimultaneousAnalysis,buhlmann2011}, guaranteeing the convergence of the high-dimensional calibration weights and the augmented outcome regression.
%The bounded error assumption in Condition \ref{as: sub gaussian of residual of outcome regression} can be relaxed to the sub-Gaussian error if the sparsity Condition \ref{as: sparsity of gamma and theta} holds with an additional factor $\log^{1/2}(n)$.
Conditions \ref{as: convergence of DR parameter gamma} and \ref{as: convergence of pretrained model} 
specify the required convergence rates of the high-dimensional density ratio and pre-trained models, which are well-established for lasso-type estimators (\cite{buhlmann2011}). 
% The convergence of the prediction error of $w_{u,k}(\bX,\widehat{\bgamma}_{u,k})$ and $w_{u,k}^{(1)}(\bX,\widehat{\bgamma}_{u,k})$ can be achieved through the convergence of parameters and Taylor expansion. 
The bounded weights condition is common for high-dimensional weighted estimators \citep{Ning2020Robust}. 
%when all the density ratio models are misspecified to guarantee the convergence of the empirical likelihood weights. 
%It's important to note that the solution of empirical likelihood corresponds to the solution of a penalized minimization problem, making this condition attainable. 
% Condition \ref{as:initial estimator of beta} is the requirement for the initial estimator in the outcome regression (\ref{eq:EL-outcome}). The estimator $\widehat{\bbeta}_{\EL,u}$ is eligible based on Theorem \ref{thm: multiply robust consistency of Z estimator}.
Condition \ref{as: estimating function} contains regularity conditions for the estimating functions \citep{PakesandDavid1989,van2000asymptotic}; its third part concerning $w_{u}^{\ast}(\bX)$ is trivial in the linear model and the GLM where ${\bS}_a$ is independent of $\bbeta$. 
%,newey2004higher
% While the third part relates to the limiting value $w_{u}^{\ast}(\bX)$ of the estimated density ratio,  it becomes trivial in the linear model and the GLM since ${\bS}_a$ is independent of $\bbeta$. 
Condition \ref{as: sparsity of gamma and theta} imposes sparsity assumptions on the nuisance parameters. The factor $\log^{1/2}(n)$ can be removed if the outcome model is correct at $\bbeta_0$. 
Compared with the assumption $\max(s_{1u},s_{2u}) \log(\widetilde{m}_u) = o(n^{1/2})$ in \citet{Tan2020Modelassited} and \citet{Ning2020Robust}, we add a sparsity component $s_{3u}$
 for estimating the pre-trained model and an extra factor $\{(s_{1u}\vee s_{3u})\log(p)\}^{1/2}$ to account for the estimation errors in the augmented basis functions $\widehat{\bfd}_u(\bX)$.
% Our sparsity condition is slightly stronger than that in \citet{Tan2020Modelassited} and \citet{Ning2020Robust}, which requires $\max(s_{1u},s_{2u}) \log(\widetilde{m}_u) = o(n^{1/2})$. 
% The additional term $ (\min(s_1, s_2))^{1/2}$ in this condition is introduced due to the feasibility of high-dimensional empirical likelihood \citep{chang2021HDEL}.
% First, we introduce an additional sparsity level $s_{3u}$ to account for the estimation of the pre-trained model $m(\bX,\widehat{\bxi}_u)$. Second, the extra term $\{(s_{1u}\vee s_{3u})\log(p)\}^{1/2}$ is needed to ensure the feasibility of the high-dimensional calibration weights due to the estimation errors in the augmented basis functions $\widehat{\bfd}_u(\bX)$.
%The sparsity level $s_{1u}$ is replaced by $\max(s_{1u},s_{3u})$ in our condition to account for the estimation error in the predicted score function $\widetilde{\bS}_1(m(\bX,\widehat{\bxi}_u),\bZ)$ in the EL formulation and outcome regression.
Including the derivatives of the pre-trained and DR models in $\widehat{\bfd}_u(\bX)$ is crucial for multiply robust inference, which is not considered in the aforementioned works.
%Furthermore, an extra term $\log^{1/2}(p)$ is included for the predicted score function, $\widetilde{\bS}_1^{(1)}(\bX;\widehat{\bxi}_u)$ in the SCB constraints, which contains an $\bX$ in its form.
% since we require the covariate balancing property of the derivative of the predicted score function, $\widetilde{\bS}_1^{(1)}(\bX;\widehat{\bxi}_u)$, which contains an $\bX$ in its form.

For two positive sequences $\{a_n\}$ and $\{b_n\}$, let $a_n \asymp b_n$ denote $c_1 \leq a_n / b_n \leq c_2$ for two positive constants $c_1, c_2$ and all $n$.
In the following, we state the main theoretical results.
% for the proposed method.

\begin{theorem}\label{thm: multiply robust consistency of Z estimator}
Under Conditions \ref{as:ignorability}--\ref{as: estimating function} for the $u$th source, $(s_{1u}\vee s_{3u})s_{\lambda u}\log(m) = o(n)$, $w_{\ps} \asymp \sqrt{ (s_{1u} \vee s_{3u})\log({m})/n}$, the high-dimensional calibration estimator $\widehat{\bbeta}_{\EL,u}$ proposed in (\ref{eq:obj}) using the $u$th source is multiply robust consistent and satisfies $|\widehat{\bbeta}_{\EL,u} - {\bbeta}_0   |_2 = O_p( \{ (s_{1u}\vee s_{3u}) \log({m}) /n\}^{1/2} )$ if only one of the DR models for the $u$th source satisfies (\ref{eq:HD-DR})  or the outcome model in (\ref{eq:HD-OR}) is correctly specified for $\bbeta \in \mathcal{B}$. 
%Moreover, $|\widehat{\bbeta}_{\EL,u} - {\bbeta}_0   |_1 = O_p( \{ (s_{1u}\vee s_{3u}) \log({m}) /n\}^{1/2} )$.
\end{theorem}

Theorem \ref{thm: multiply robust consistency of Z estimator} establishes the multiply robust consistency of the proposed estimator $\widehat{\bbeta}_{\EL,u}$, with a slower convergence rate than $n^{-1/2}$ due to nuisance-parameter estimation. This shows the necessity of debiasing the estimator.
% using the data from the $u$th source that satisfies the transferability and regularity conditions. 
% However, the convergence rate is slower than $n^{-1/2}$, due to the high dimensionality of the DR and OR models and the soft balancing in (\ref{eq:EL-CB}). 
% This shows the necessity of debiasing the estimator.
%due to the convergence rate of the estimated loss function being of order $O_p(\sqrt{\frac{(s_{1u}\vee s_{3u})\log(\widetilde{m})}{n}})$. 

Theorem S1 in the SM 
% \ref{thm: estimating eqs convergence}  
shows that the debiasing procedure in Section \ref{sec: single source inference} leads to the asymptotically unbiased estimating equations for the target population $\mathbb{Q}$. 
Now we present the asymptotic property of the estimator $\widehat{\bbeta}_{\AEL,u}$ in (\ref{eq:obj-HD}) using data from a single source.

Let $\psi_{u, 1, j}(Y_i,\bX_i;\bbeta) = w_{u,i}^{\ast} \{{S}_j(Y_i, \bZ_i; \bbeta) - \psi_{u, 2, j}(\bX_i;\bbeta)\}$ and $\psi_{u, 2, j}(\bX_i;\bbeta) ={S}_{bj}(\bZ,\bbeta)\allowbreak  +  
\btheta_{u,j}^{\ast\T}\bU_{u,i}^{\ast}$.
% {\btheta}_{u,j,1}^{\ast \T} {\bfd}_u^{\ast}(\bX_i) + {\btheta}_{u,j,2}^{\ast \T} {\ba}_{u,1i}^{\ast} + {\btheta}_{u,j,3}^{\ast \T} {\ba}_{u,2i}^{\ast}$. 
Recall that $\psi_{u, 1, j}(Y_i,\bX_i) = \psi_{u, 1, j}(Y_i,\bX_i;\bbeta_0)$, $\psi_{u, 2, j}(\bX_i) = \psi_{u, 2, j}(\bX_i;\bbeta_0)$ and  $\bV_{0,u} = {\E}_{\mathbb{F}_u} \{ w_u^{\ast}(\bX) \frac{\partial}{\partial \bbeta} {\bS}_a(Y, \bZ; \bbeta_0)\} + {\E}_{\mathbb{Q}} \{ \frac{\partial}{\partial \bbeta} {\bS}_b(\bZ, \bbeta_0)\}$, defined after (\ref{eq:V0}) in Section \ref{sec:test}. It can be inferred from Theorem S1 that under a correctly specified DR model, $\bV_{0,u} = {\E}_{\mathbb{Q}} \{ \frac{\partial}{\partial \bbeta} \bS(Y, \bZ; \bbeta_0)\}$, but this may not hold if all the DR models are misspecified.

\begin{theorem}\label{thm: IFs of target parameter}
Under Conditions \ref{as:ignorability}--\ref{as: sparsity of gamma and theta}, 
$\omega_{\ps} \asymp \{(s_{1u}\vee s_{3u})\log(\widetilde{m}_u)\log^{1/2}(p)/n \}^{1/2}$, $\omega_{\oc} \asymp \{\log(\widetilde{m}_u)/n \}^{1/2}$, 
giving an initial estimator $\widehat{\bbeta}^{\ini}$ satisfying $(s_{1u}\vee s_{3u}) \log(\widetilde{m}_u)|\widehat{\bbeta}^{\ini} - \bbeta_0 |_2^2  = o_p(1)$, 
% and $|\widehat{\bbeta}^{\ini} - \bbeta_0 |_1 = O_p( \{ (s_{1u}\vee s_{3u}) \log(\widetilde{m}_u) /n\}^{1/2} )$ for the initial estimator,
if only one of the DR models for the $u$th source satisfies (\ref{eq:HD-DR}) or the outcome model in (\ref{eq:HD-OR}) at $\bbeta_0$ is correctly specified, 
then we have $\widehat{\bbeta}_{\AEL,u} - \bbeta_0 = o_p(1)$ and
\be
\sqrt{n_0}(\widehat{\bbeta}_{\AEL,u} - \bbeta_0) = - n_0^{-1/2} \bV_{0,u}^{-1} \bigg\{\sum_{i\in S_{\mathbb{F}_u }} c_u^{-1} \bpsi_{u,1}(Y_i,\bX_i) + \sum_{i \in S_{\mathbb{Q}}} \bpsi_{u,2}(\bX_i) \bigg\} + o_p(1), \nn 
\ee
where $\bpsi_{u,\ell}(Y_i,\bX_i) = (\psi_{u,\ell,1}(Y_i,\bX_i),\ldots, \psi_{u,\ell,q}(Y_i,\bX_i))^{\T}$ for $\ell = 1,2$ and $c_u = \lim n_{1u}/n_0$.
		% \bea
		% \psi(Y_i,\bX_i,D_i;\bbeta)(j) &=& \frac{D_i}{1-W}w(\bX_i)\{S_j(Y_i, \bZ_i; \bbeta) - {\btheta}_{j,1}^{\ast \T} {\bfd}^{\ast}(\bX_i) - {\btheta}_{j,2}^{\ast \T} {\ba}_{1i}^{\ast} - {\btheta}_{j,3}^{\ast \T} {\ba}_{2i}^{\ast}\} + \nn \\
		% &&\frac{1-D_i}{W} \{ {\btheta}_{j,1}^{\ast\T} {\bfd}^{\ast}(\bX_i) + {\btheta}_{j,2}^{\ast\T} {\ba}_{1i}^{\ast} + {\btheta}_{j,3}^{\ast\T} {\ba}_{2i}^{\ast} \}  \nn
		% \eea 
		% and $V_{\bbeta_0}$ is the variance matrix evaluated at the true parameter $\bbeta_0$, i.e. $V_{\bbeta_0} = \E\{ \frac{\partial}{\partial \bbeta}\bS(Y,\bX;\bbeta) \big|_{\bbeta = \bbeta_0} \} $.
\end{theorem}

%Note that, given an initial consistent estimator, this theorem does not require the correctness of the outcome model in (\ref{eq:HD-OR}) for all values of $\bbeta$ in the parameter space.

The tuning parameter $w_{\ps}$ includes an extra term $(s_{1u}\vee s_{3u})^{1/2}\log^{1/4}(p)$ due to the estimation errors 
% in the augmented basis functions $\widehat{\bfd}_u(\bX)$ introduced by estimating 
in the DR and pre-trained models. The influence functions have the same form under either a correctly specified DR model, where $w_{u, i}^{\ast} = w_u(\bX_i)$, or a correctly specified OR model, where $ \E_{\mathbb{Q}}\{{S}_{aj}(Y,\bZ; \bbeta_0)\vert \bX = \bX_i\} = 
\btheta_{u,j}^{\ast \T}\bU_{u,i}^{\ast}$
% {\btheta}_{u,j,1}^{\ast \T} {\bfd}_u^{\ast}(\bX_i) + {\btheta}_{u,j,2}^{\ast \T} {\ba}_{u,1i}^{\ast} + {\btheta}_{u,j,3}^{\ast \T} {\ba}_{u,2i}^{\ast}$ 
for all $j$, achieving multiply robust inference. 
Any initial estimator with $(s_{1u}\vee s_{3u}) \log(\widetilde{m}_u)|\widehat{\bbeta}^{\ini} - \bbeta_0 |_2^2  = o_p(1) $
% = o_p( \{ (s_{1u}\vee s_{3u}) \log(\widetilde{m}_u) /n\}^{1/2} )$ 
can be employed in the outcome regression (\ref{eq:AEL-g}), and $\widehat{\bbeta}_{\EL,u}$ serves as an eligible candidate by Theorem \ref{thm: multiply robust consistency of Z estimator}.
% The requirement of the convergence rate for the initial estimator is needed in the augmented outcome regression in (\ref{eq:EL-outcome}). From Theorem \ref{thm: multiply robust consistency of Z estimator}, the EL estimator $\widehat{\bbeta}_{\EL,u}$ satisfies this condition and is eligible for the initial estimator.
%Furthermore, if one of the density ratio models is correctly specified, $w_{u, i}^{\ast} = w_u(\bX_i)$, the true density ratio; if the outcome model is correctly specified, $ \psi_{u, 2, j}(Y_i,\bX_i;\bbeta) =  \widetilde{S}_{j2}(\bZ_i,\bbeta) + \E_{\mathbb{Q}}\{\widetilde{S}_j(Y,\bZ; \bbeta_0)\vert \bX = \bX_i  \}  $. The convergence rate $o_p(n^{-1/2})$ is necessary for the inference of the estimator from equation (\ref{eq:obj-HD}). While Theorem \ref{thm: multiply robust consistency of Z estimator} requires the outcome model to be correctly specified for all possible values of $\bbeta$, 
Notably, with a consistent initial estimator, Theorem \ref{thm: IFs of target parameter} only requires the OR model to be correctly specified at the true parameter value $\bbeta_0$.

Theorem \ref{thm: IFs of target parameter} establishes the influence functions for the proposed debiased estimator $\widehat{\bbeta}_{\AEL,u}$ from the estimating equations using the $u$th source data, implying the asymptotic normality 
$$\sqrt{n_0}(\widehat{\bbeta}_{\AEL,u} - \bbeta_0) \overset{d}{\to} \mathcal{N}(\mathbf{0}_{q}, \bSigma_{\AEL}^{(u u)}),$$ 
where 
$\bSigma_{\AEL}^{(u u)} = \bV_{0,u}^{-1} \big[ c_u^{-1}{\E}_{\mathbb{F}_u}\{\bpsi_{u, 1}(Y,\bX)\bpsi_{u, 1}^{\T}(Y,\bX)\} + {\E}_{\mathbb{Q}}\{\bpsi_{u, 2}(\bX)\bpsi_{u, 2}^{\T}(\bX)\} \big] \bV_{0,u}^{-1}$ is defined in (\ref{eq:cov-1}).
Let $\widehat{\psi}_{u, 1, j} = n_{1u}\widetilde{p}_{u,i} \{{S}_{aj}(Y_i, \bZ_i; \widehat{\bbeta}_{\AEL,u}) - \widehat{\psi}_{u, 2, j}\}$, $\widehat{\psi}_{u, 2, j} = {S}_{bj}(\bZ_i,\widehat{\bbeta}_{\AEL,u}) + 
\widetilde{\btheta}_{u,j}^{\T}\widehat{\bU}_{u,i} $,
% \widetilde{\btheta}_{u,j,1}^{ \T} \widehat{\bfd}_u(\bX_i) + \widetilde{\btheta}_{u,j,2}^{ \T} \widehat{\ba}_{u,1i} + \widetilde{\btheta}_{u,j,3}^{ \T} \widehat{\ba}_{u,2i}$, 
and $\widehat{\bpsi}_{u,k} =(\widehat{\psi}_{u,k,1},\ldots,\widehat{\psi}_{u,k,q})^{\T}$ for $k = 1,2$. The estimate of $\bSigma_{\AEL}^{(u u)}$ can be constructed as
% $\widehat{\bSigma}_{\AEL}^{(u u)} =\widehat{\bV}_0^{-1}\widehat{\bOmega}_u\widehat{\bV}_0^{-1}$ where 
    % \bea
    % \widehat{\Sigma}_{\AEL} &=& \widehat{\bV}_0^{-1}\widehat{\bOmega}\widehat{\bV}_0^{-1} \nn \\ 
    % % \widehat{\bV}_0 &=&  \nn \\ 
    % \widehat{\bOmega} &=& \frac{1}{n_1}\sum_{i\in S_{\mathbb{F}}} \widehat{\psi}_1(Y_i,\bX_i;\widehat{\bbeta}_{\AEL})\widehat{\psi}_1^{\T}(Y_i,\bX_i;\widehat{\bbeta}_{\AEL}) + \frac{1}{n_0}\sum_{i \in S_{\mathbb{Q}}}\widehat{\psi}_2(Y_i,\bX_i;\widehat{\bbeta}_{\AEL})\widehat{\psi}_2^{\T}(Y_i,\bX_i;\widehat{\bbeta}_{\AEL}) \nn 
    % \eea
    \bea
    \widehat{\bSigma}_{\AEL}^{(u u)} &=& \widehat{\bV}_{0,u}^{-1} \bigg( \frac{n_0}{n_{1u}^2}\sum_{i\in S_{\mathbb{F}_u}} \widehat{\bpsi}_{u,1}\widehat{\bpsi}_{u,1}^{\T} + \frac{1}{n_0}\sum_{i \in S_{\mathbb{Q}}}\widehat{\bpsi}_{u,2}\widehat{\bpsi}_{u,2}^{\T} \bigg) \widehat{\bV}_{0,u}^{-\T} \mbox{ \ where } \label{eq: Sigma_uu estimate} \\ 
    % \widehat{\bOmega}_u &=& \frac{n_0}{n_{1u}^2}\sum_{i\in S_{\mathbb{F}_u}} \widehat{\bpsi}_{u,1}\widehat{\bpsi}_{u,1}^{\T} + \frac{1}{n_0}\sum_{i \in S_{\mathbb{Q}}}\widehat{\bpsi}_{u,2}\widehat{\bpsi}_{u,2}^{\T} \mbox{ \ and } \nn \\
    \widehat{\bV}_{0,u} &=& \sum_{i \in S_{\mathbb{F}_u}} \widetilde{p}_{u,i}\frac{\partial}{\partial \bbeta} {\bS}_a(Y_i,\bZ_i;\widehat{\bbeta}_{\AEL,u}) + \frac{1}{n_0}\sum_{i \in S_{\mathbb{Q}}} \frac{\partial}{\partial \bbeta} {\bS}_b(\bZ_i,\widehat{\bbeta}_{\AEL,u}). \nn
    \eea
% The formulation of $\widehat{\bV}_{0,u}$ depends on the estimating function. 
Particularly, for estimating equations in the linear model and the GLM, the estimator of $\bV_{0,u}$ can be simplified. If $\bS(Y,\bZ; \bbeta) = (Y - \bZ^{\T} \bbeta)\bZ$, it suffices to use $\widehat{\bV}_{0,u} = n_0^{-1}\sum_{i\in S_{\mathbb{Q}}}\bZ_i\bZ_i^{\T}$.
% as $\frac{\partial}{\partial \bbeta} {\bS}_a(Y,\bZ;\bbeta) = \bzero$. 

% one may consider the first-order difference for an approximation, as suggested by Theorem 7.4 of \citet{Newey&McFadden1994largesample}.

% The estimate $\widehat{\bSigma}_{\AEL}^{(u_1 u_2)}$ of $\bSigma_{\AEL}^{(u_1 u_2)}$ in (\ref{eq:cov-2}) can be similarly constructed.

Based on Theorem \ref{thm: IFs of target parameter}, the following proposition validates the proposed specification test for the transferability of all sources presented in Section \ref{sec:test}.
% Theorem \ref{thm: IFs of target parameter} reveals that, if one of the density ratio models is correctly specified for each source, and the outcome model is correct, the proposed estimator attains the semi-parametric efficiency bound established in \citet{chen2008semi}.

\begin{proposition}
\label{prop: specification test}
    % Under the null hypothesis that Condition \ref{as:ignorability} holds for all sources,
    Suppose the conditions of Theorem \ref{thm: IFs of target parameter} hold for all sources, which implies the null hypothesis of (\ref{eq:Hypotheses}). The test statistic 
    \be
    F = n_0 (\bC\widehat{\bbeta}_{\AEL})^{\T}(\bC\widehat{\bSigma}_{\AEL}\bC^{\T}  )^{-1}(\bC\widehat{\bbeta}_{\AEL}) \nn
    \ee
    is asymptotically $\chi^2_{(g-1)q}$ distributed with $(g-1)q$ degrees of freedom, where $\widehat{\bSigma}_{\AEL} = (\widehat{\bSigma}_{\AEL}^{(u_1 u_2)})$.
    % where $\widehat{\Sigma}_{\AEL}$ is the estimate of the covariance $\Sigma_{\AEL}$, obtained by plugging in the estimates $\widehat{\bV}_0$, $\widehat{\psi}_{u, 1}(Y,\bX)$ and $\widehat{\psi}_{u, 2}(Y,\bX)$ into $\bSigma_{\AEL}^{(u_1 u_2)}$. 
\end{proposition}

% The proposed specification test rejects the null hypothesis in (\ref{eq:Hypotheses}) if $F > \chi^2_{(g-1)q}(\alpha)$.
The estimator $\widehat{\bSigma}_{\AEL}^{(u_1 u_2)}$ 
% of $\bSigma_{\AEL}^{(u_1 u_2)}$ 
can be constructed similarly to (\ref{eq: Sigma_uu estimate}).
Under the null hypothesis, 
Proposition \ref{prop: specification test} guarantees that the proposed test controls the asymptotic Type I error. 
% at level $\alpha$.
% establishes the asymptotic control of the Type I error at $\alpha$ of the proposed test.
% demonstrates that the proposed test can control the Type I error rate at $\alpha$ asymptotically under the null hypothesis. 

In the following, we derive the asymptotic properties and the efficiency of the multi-source debiased estimator $\widehat{\bbeta}_{\MDEL}$.
% , which combines information from all transferable sources. 
Recall that $w_i^{\ast}$ is the probability limit of the estimated density ratio from equations  (\ref{eq:EL-observe-multi})-(\ref{eq:EL-CB-HD-multi-source}).
Let $\psi_{ 1, j}(Y_i,\bX_i) = w_{i}^{\ast} \{{S}_j(Y_i, \bZ_i; \bbeta_0) - \psi_{ 2, j}(\bX_i)\}$, $\psi_{2, j}(\bX_i) = {S}_{bj}(\bZ_i,\bbeta_0) +  \btheta_{j}^{\ast \T}\bU_i^{\ast}$,
% {\btheta}_{j,1}^{\ast \T} {\bfd}^{\ast}(\bX_i) + {\btheta}_{j,2}^{\ast \T} {\ba}_{1i}^{\ast} + {\btheta}_{j,3}^{\ast \T} {\ba}_{2i}^{\ast}$, 
$\bpsi_k = (\psi_{k,1},\ldots ,\psi_{k,q})^{\T}$ for $k = 1,2$
% $\bpsi_{1}(Y_i,\bX_i) = (\psi_{1,1}(Y_i,\bX_i),\ldots, \psi_{1,q}(Y_i,\bX_i))^{\T}$, $\bpsi_{2}(\bX_i) = (\psi_{2,1}(\bX_i),\ldots,\psi_{2,q}(\bX_i))^{\T}$ 
and $\tilde{c}_g = \sum_{u = 1}^{g} c_u = \lim n_{1}/n_0$. 
Similar to $\blambda_{u,h}^{\ast}$, let $\blambda^{\ast}_{h}$, $h = 1, 2, 3,4$, be the probability limits of the estimated parameters in the conjugate optimization of the multi-source calibration in (\ref{eq:EL-observe-multi}).
Let $s_{1} = \max_{u\in [g]} s_{1u} $, $s_2 = \max_{j \in [q]} |(\btheta_{j,1}^{\ast \T},\btheta_{j,2}^{\ast \T},\btheta_{j,3}^{\ast \T}) |_0 $, $s_{3} = \max_{u\in [g]} s_{3u} $ and $s_{\lambda} = |\blambda_{2}^{\ast}|_0$. 

Under the multi-source setting, the expression for the correctly specified DR model and OR model can be formulated for some coefficients $\bgamma_{u,k}^{(0)}$, $\bxi_u^{(0)} $, and basis functions $\bb(\bx)$ as
\bea
\frac{1}{w(\bx)} = \frac{d \mathbb{F}_{\bX}}{d \mathbb{Q}_{\bX}} & \in & \operatorname{span}\{w_{u,k}^{-1}(\bx,\bgamma_{u,k}^{(0)}), u \in [g] , k \in [v_u] \}, \mbox{\ and} \label{eq: multi source DR model} \\
{\E}_{\mathbb{Q}} \{{S}_{aj}(Y, \bZ; \bbeta_0) \mid \bX = \bx\}  &\in & \operatorname{span}\{\widetilde{S}_{aj}(m(\bx, \bxi_u^{(0)}), \bZ), u\in [g], b_1(\bx), \ldots, b_{m}(\bx)\}.
\label{eq: HD-OR-multisource} 
\eea

\begin{theorem}\label{thm: IFs of target parameter Multi Source}
% Suppose that the conditions in Theorem \ref{thm: IFs of target parameter} hold for all sources and the sparsity levels of the parameters are defined as the maximum sparsity across all sources, $s_{s} = \underset{u\in [q]}{\sup} s_{su} $ for $s = 1,3$ and $\lambda$, and $s_2 = \underset{j \in [q]}{\sup} |(\btheta_{j,1}^{\ast \T},\btheta_{j,2}^{\ast \T},\btheta_{j,3}^{\ast \T}) |_0 $ . 
Suppose Conditions \ref{as:ignorability} and \ref{as: convergence of pretrained model} hold for all sources, and Conditions \ref{as: sub gaussian of basis functions and covariates}--\ref{as: convergence of DR parameter gamma} and \ref{as: estimating function} hold for the mixture source population $\mathbb{F}$ and the multi-source calibration weights $w_{i}^{\ast}$,
$\sqrt{s_{1}\vee s_{3}}(s_{1}\vee s_{2} \vee s_{3})\log(\widetilde{m})\log^{1/2}(p)\log^{1/2}(n)= o(n^{1/2})$, $(s_{1}\vee s_{3})s_{\lambda}\log(\widetilde{m}) = o(n)$, $\omega_{\ps} \asymp \{(s_{1}\vee s_{3})\log(\widetilde{m})\log^{1/2}(p)/n \}^{1/2}$, $\omega_{\oc} \asymp \{\log(\widetilde{m})/n\}^{1/2}$,
% and  $|\widehat{\bbeta}^{\ini} - \bbeta_0 |_1 = O_p( \{ (s_{1}\vee s_{3}) \log(\widetilde{m}) /n\}^{1/2} )$, 
giving an initial estimator $\widehat{\bbeta}^{\ini}$ satisfying $(s_{1}\vee s_{3}) \log(\widetilde{m})|\widehat{\bbeta}^{\ini} - \bbeta_0 |_2^2  = o_p(1)$, 
if either (\ref{eq: multi source DR model}) or (\ref{eq: HD-OR-multisource}) holds for a unique set of coefficients respectively, the MHDC estimator $\widehat{\bbeta}_{\MDEL}$ satisfies $$\sqrt{n_0}(\widehat{\bbeta}_{\MDEL} - \bbeta_0) = - n_0^{-1/2} \bV_{0}^{-1} \bigg\{ \sum_{i\in S_{\mathbb{F} }} \tilde{c}_g^{-1} \bpsi_{1}(Y_i,\bX_i) + \sum_{i \in S_{\mathbb{Q}}} \bpsi_{2}(\bX_i) \bigg\} + o_p(1),$$
        % $\bpsi_{1} = (\psi_{1,1},\ldots, \psi_{1,q})^{\T}$, $\bpsi_{2} = (\psi_{2,1},\ldots,\psi_{2,q})^{\T}$, and $\psi_{1,j},\psi_{2,j}$ are defined in the same way as $\psi_{u,1,j}$ and $\psi_{u,2,j}$.
where $ \bV_{0} = \E_{\mathbb{F}} \{w^{\ast}(\bX)\frac{\partial}{\partial\bbeta} {\bS}_a(Y,\bZ;\bbeta) \mid_{\bbeta = \bbeta_0}  \} + \E_{\mathbb{Q}} \{ \frac{\partial}{\partial \bbeta}{\bS}_b(\bZ,\bbeta)  \mid_{\bbeta = \bbeta_0}  \} $.
\end{theorem}

Theorem \ref{thm: IFs of target parameter Multi Source} derives the influence function of the proposed MHDC estimator $\widehat{\bbeta}_{\MDEL}$ and leads to the asymptotic normality result $\sqrt{n_0}(\widehat{\bbeta}_{\MDEL} - \bbeta_0) \overset{d}{\to} \mathcal{N}(\mathbf{0}_{q}, \bSigma_{\MDEL} )$, where 
\be\label{eq:sigma-multi}
\bSigma_{\MDEL} = \bV_0^{-1} \big[ \tilde{c}_g^{-1}{\E}_{\mathbb{F}}\{\bpsi_{1}(Y,\bX)\bpsi_{1}^{\T}(Y,\bX)\} + {\E}_{\mathbb{Q}}\{\bpsi_{2}(\bX)\bpsi_{2}^{\T}(\bX)\} \big] \bV_0^{-1}.
\ee
An asymptotic $1-\alpha$ confidence interval of $\beta_{0,j} $ can be constructed as 
\be
\text{CI}_{j,\alpha} = (\widehat{\beta}_{\MDEL,j} - z_{1-\alpha/2}(\widehat{\sigma}_j/n_0)^{1/2},  \widehat{\beta}_{\MDEL,j} + z_{1-\alpha/2}(\widehat{\sigma}_j/n_0)^{1/2} ), 
\label{eq: confidence interval}
\ee
where $z_{1-\alpha/2}$ is the upper $\alpha/2$ quantile of the standard normal distribution, $\widehat{\sigma}_j = \bfe_j^{\T} \widehat{\bSigma}_{\MDEL}\bfe_j$, $\widehat{\bSigma}_{\MDEL}$ is the estimate of $\bSigma_{\MDEL}$, constructed in the same way as
% by plugging in the estimated $\widehat{\bV}_0$, $\widehat{\bpsi}_1(Y,\bX)$ and $\widehat{\bpsi}_{2}(\bX)$ 
$\widehat{\bSigma}_{\AEL}^{(u u)}$ in (\ref{eq: Sigma_uu estimate}), and $\bfe_j$ is the unit vector with the $j$th element being $1$.

Compared to the single-source estimator $\widehat{\bbeta}_{\AEL,u}$ in Theorem \ref{thm: IFs of target parameter}, the first part of the influence function, $\bpsi_{1}(Y_i,\bX_i)$, of the multi-source estimator $\widehat{\bbeta}_{\MDEL}$ combines the data from all sources $S_{\mathbb{F}}$, improving the efficiency of $\widehat{\bbeta}_{\MDEL}$.

Recall that $\widehat{\bbeta}_{\A} = \sum_{u=1}^g \bA_u \widehat{\bbeta}_{\AEL, u}$ is the linear combination of single-source estimators $\{\widehat{\bbeta}_{\AEL, u}\}$ with $\sum_{u=1}^g \bA_u = \bI_q$. Note that $n_0 \var(\widehat{\bbeta}_{\A}) = \bA \bSigma_{\AEL} \bA^{\T} \{1 + o(1)\}$, where $\bA = (\bA_1, \ldots, \bA_g) \in \mathbb{R}^{q\times qg}$. Let $\bM \succ 0$ denote a positive definite matrix $\bM$.
The next theorem compares $\bSigma_{\MDEL}$ and $\bA \bSigma_{\AEL} \bA^{\T}$ for the asymptotic variances of  $\widehat{\bbeta}_{\MDEL}$ and $\widehat{\bbeta}_{\A}$.  

% Last, we present a proposition that shows the efficiency gain in combining all sources, that the asymptotic variance is no larger than the averaged estimators from separate sources.

\begin{theorem}
\label{thm: efficiency gain of combining all sources}
Suppose that the conditions in Theorem \ref{thm: IFs of target parameter Multi Source} hold. If for each source, the OR model and one of the DR models are correctly specified, then there is an efficiency gain such that 
%For each $u \in [g]$, the estimator $\widehat{\bbeta}_{\AEL,u}$ is calculated from (\ref{eq:obj-HD}) using combined data $S_{\mathbb{F},u}\cup S_{\mathbb{Q}}$, and $\widehat{\bbeta}_{\AEL}$ is calculated from (\ref{eq:obj-HD-multi}) combining all sources. Let ${\widehat{\bbeta}}_{\AEL,\bA}$ be the averaged estimator with weight matrices $\bA = (\bA_1,\ldots,\bA_g) \in  \mathbb{R}^{q\times qg}$
% $\sum_{u=1}^g A_u = I_q $
$\bA {\bSigma}_{\AEL} \bA^{\T} - \bSigma_{\MDEL} \succ 0$
for any $\bA = (\bA_1, \ldots, \bA_g) \in \mathbb{R}^{q\times qg}$ satisfying $\sum_{u=1}^g \bA_u = \bI_q$. Particularly, $\bA^{\ast} {\bSigma}_{\AEL} \bA^{\ast \T} - \bSigma_{\MDEL} \succ 0$, where $\bA^{\ast}$ is the optimal weighting matrix in (\ref{eq:opt-weight}).
\end{theorem}

Theorem \ref{thm: efficiency gain of combining all sources} demonstrates that the MHDC estimator is strictly more efficient than any linear combination estimator $\widehat{\bbeta}_{\A}$ asymptotically, including the optimally weighted estimator $\widehat{\bbeta}_{\A, \opt}$ in (\ref{eq:combined-est}). This indicates the superiority of using high-dimensional calibration to combine information from all transferable sources.

%%%%%%%%%%%%%%%%%%%%%%%%%%%%%%%%%%%%%%%%%%%%%
%%%%%%%%%%%%%%%%%%%%%%%%%%%%%%%%%%%%%%%%%%%%%

%\setcounter{equation}{0}
%\section{Adaptive learning}\label{sec:adaptive}

%%%%%%%%%%%%%%%%%%%%%%%%%%%%%%%%%%%%%%%%%%%%%
%%%%%%%%%%%%%%%%%%%%%%%%%%%%%%%%%%%%%%%%%%%%%

\setcounter{equation}{0}
\section{Simulation study}\label{sec:simulation}

In this section, we evaluate the performance of the proposed multi-source estimator $\widehat{\bbeta}_{\MDEL}$ (MHDC)
% in (\ref{eq:obj-HD-multi}) 
and compare it with several alternatives, including the HD calibration (HC) estimator using weights $\{\widetilde{p}_i\}$ from (\ref{eq:EL-observe-multi}) without the constraint (\ref{eq:EL-CB-projection-multi}); the linear combination of the single-source HDC estimators (LC-HDC) in  (\ref{eq:combined-est}),
% the high-dimensional semi-supervised estimator (HDSS) in \citet{Deng2024optimal}, 
and the doubly robust AIPW estimator (DR-AIPW), modified from \citet{Zhou2024doubly} for general estimating equations. 
All methods use the full source data for a fair comparison.
% The last two methods are applied to the combined source data, so that those five methods use all the source data for a fair comparison.

Let $\bX = (X_1,\ldots,X_p)^{\T}$ denote the covariates. We consider $\mathcal{N}(\bmu_{\mathbb{Q}},\bSigma_{\mathbb{Q}})$ and $\mathcal{N}(\bmu_u,\bSigma_u)$ for the target and source distributions of $\bX$, where we take $\bmu_{\mathbb{Q}} \neq \bmu_u$ or $\bSigma_{\mathbb{Q}} \neq \bSigma_u$ to simulate the cases of covariate shift. 
Multiple sources are obtained by varying $\bmu_u$ and $\bSigma_u$. Let $\bSigma_{\phi,p} = (\sigma_{\phi, j_1j_2})_{p \times p}$ be a $p$-dimensional $\AR(1)$ covariance matrix with parameter $\phi$, where $\sigma_{j_1j_2} = \phi^{|j_1-j_2|}$ for $1 \leq j_1, j_2 \leq p$. The target distribution is fixed with $\mu_{\mathbb{Q}} = \bzero_p$ and $\bSigma_{\mathbb{Q}} = \bSigma_{0.3,p}$. Let $w_u(\bX,\bgamma_u) = \frac{d\mathbb{Q}}{d\mathbb{F}_u}(\bX)$ be the density ratio, where $\bgamma_u$ depends on $\bmu_u$ and $\bSigma_u$. We consider three scenarios of covariate shift with two source distributions.

% on the source and target domains, generated from Gaussian distributions $\mathcal{N}(\bmu_u,\bSigma_u)$ and , respectively.  Subscripts $u$ and $\mathbb{Q}$ indicate different source domains and the target domain to highlight the pattern of covariate shift. Let $\bSigma_{\phi,p} = (\sigma_{j_1j_2})$ be the covariance matrix of an $\AR(1)$ model with $\sigma_{j_1j_2} = (\phi^{|j_1-j_2|} - \phi^{j_1+j_2})/(1-\phi^2) $

% where $U_j = \sqrt{\phi} U_{j-1} + \sqrt{1 - \phi}\omega_j$ for $j \in [p]$ with $U_0 = 0$ and $\omega_j \overset{\IID}{\sim} \mathcal{N}(0,1)$. 
% We consider three different covariate shift scenarios and the density ratios $w_u(\bX,\bgamma_u) = \frac{d\mathbb{Q}}{d\mathbb{F}_u}(\bX)$, where $\bgamma_u$ depends on the shifting parameters $\mu$ and $\phi$: 
\begin{itemize}
    \item {\bf{Mean shift}}: $\bmu_{u} = (\mu\bone_{10}^{\T},\bzero_{p-10}^{\T})^{\T} $ and $\bSigma_{u} = \bSigma_{\mathbb{Q}} $ for $\mu = \pm 0.125$; the corresponding DR takes the form $w_{u}(\bX,\bgamma_{u}^{(0)}) = 
    \exp(\sum_{j=1}^{11} \gamma_{u,j}^{(0)}X_j)$, where $\gamma_{u,1}^{(0)} = \mu(1-\phi)/(1-\phi^2)$, $\gamma_{u,j}^{(0)} = \mu(1+\phi^2 - 2\phi)/(1-\phi^2)$ for $j = 2, \ldots, 9$ and $\gamma_{u,11}^{(0)} = -\mu\phi/(1-\phi^2)$.
    
    \item {\bf{Cov shift}}: $\bmu_{u} = \bmu_{\mathbb{Q}}$ and $\bSigma_{u} = \diag(\bSigma_{\phi,10},\bSigma_{0.3,p-10} ) $ for $\phi = 0.1, 0.5$; the DR takes the form 
    $w_{u}(\bX,\bgamma_{u}^{(0)}) = 
    \exp\{\bgamma_{u,0}^{(0)} + 
    \sum_{j=1}^{10} (\gamma_{u,j\, j}^{(0)}X_j^2 + \gamma_{u,j\, j+1}^{(0)}X_jX_{j+1})  \} $.

    \item {\bf{Mean-Cov shift}}:  $\bmu_{u} = (\mu\bone_{10}^{\T},\bzero_{p-10}^{\T})^{\T}$ and $\bSigma_{u} = \diag(\bSigma_{\phi,10},\bSigma_{0.3,p-10} )$ for $(\mu,\phi) = (0.125,0.5)$ and $(\mu,\phi) = (-0.125,0.5)$; the DR takes the form 
    $w_{u}(\bX,\bgamma_{u}^{(0)}) = 
    \exp\{ \bgamma_{u,0}^{(0)} + \sum_{j=1}^{11} \gamma_{u,j}^{(0)}X_j + \sum_{j=1}^{10} (\gamma_{u,j \, j}^{(0)}X_j^2 + \gamma_{u,j \, j+1}^{(0)}X_jX_{j+1}) \}$.
\end{itemize}
The expression of the true DR parameter $\bgamma_{u}^{(0)}$ for the Cov shift and Mean-Cov shift cases can be similarly obtained. But the expressions are more involved, which are omitted here.

% For source populations, t
The response $Y$ is generated as $
Y = 3.14 + \bX^{\T} {\boldsymbol{ \varpi}} + \epsilon$, where ${\boldsymbol{ \varpi}}  = (1.37,0.98,0.78,1.28,\allowbreak 1.73,0.68,0.29,1.29,0.54,1.80,\bzero_{p-10}^{\T})^{\T}$ and $\epsilon {\sim} \mathcal{N}(0,1)$ is independent of $\bX $. Let $\mathbb{F}_{1}$ and $\mathbb{F}_{2}$ denote the two source populations,
% under each shift scenario
and $\mathbb{Q}$ denote the target population with $\mathbb{Q}_{Y \mid \bX} = \mathbb{F}_{u, Y \mid \bX}$. We generate $\{\bX_{u, i}, Y_i\}_{i = 1}^{n_1}$ i.i.d.\,from $\mathbb{F}_{u}$ for $u = 1, 2$, and $\{\bX_i\}_{i = 1}^{n_0}$ i.i.d.\,from $\mathbb{Q}$.
We consider three sets of target covariates: $\bZ_{(1)} = X_3$, $\bZ_{(2)} = (X_1, X_2, X_3)$, and $\bZ_{(3)} = (X_1, \ldots, X_{10})$, and the parameter of interest $\bbeta_{k,0}$ on $\mathbb{Q}$ is defined by $\E_{\mathbb{Q}} (Y - \bZ_{(k)}^\T\bbeta_{k,0})\bZ_{(k)} = \bzero $ for $k = 1, 2, 3$. For linear regression, the decomposition (\ref{eq:3-1}) is satisfied with $\bS_a(Y,\bZ_{(k)};\bbeta) = Y\bZ_{(k)}  $ and $\bS_b(\bZ_{(k)},\bbeta) = -(\bZ_{(k)}^{\T}\bbeta)\bZ_{(k)}$. We set the sample sizes $(n_0, n_1) = (40, 80)$ and $(n_0, n_1) = (100, 200)$, and the dimension $p = 200, 500$. Each scenario is repeated 500 times.

The procedure uses three DR models for each source: the correctly specified model $w_{u}(\bX,\bgamma_{u})$ derived for each shift scenario with two misspecified working models
$w_{u,1}(\bX,\bgamma_{u,1})\allowbreak = \exp(\bgamma_{u,1,0} + 
\sum_{j=1}^{10} \gamma_{u,1,j}X_{j+5})$, $
w_{u,2}(\bX,\bgamma_{u,2}) = \exp(\bgamma_{u,2,0} + 
\sum_{j=1}^{10} \gamma_{u,2,j\, j+1}X_{j}X_{j+1})$.
% First, we introduce the single-source case where the covariates are generated from a mixed distribution  $\Tilde{\bX} \sim (1-W)\mathbb{F}_{\bX} + W\mathbb{Q}_{\bX}$. Then Bayes's theorem is applied to calculate $\PP(D = 1 \vert \bX)$ and the indicator $D_i$ is generated accordingly as a binomial distribution. 
% In the first three cases, 
Given a pre-trained model $m(\bX,\bxi_u)$ for $\E_{\mathbb{Q}}(Y \vert \bX)$, we set $\widetilde{\bS}_{a}(m(\bX,\bxi_u), \bZ_{(k)}) = m(\bX,\bxi_u) \bZ_{(k)}$ in (\ref{eq:HD-OR}). 
Two configurations of the pre-trained model and high-dimensional covariate functions $\bb(\bX)$ are considered. The first adopts $\bb(\bX) = (1, X_1, \ldots, X_p, \{X_iX_j\}_{i,j\leq 10})$ and $m(\bX,\bxi_u) = \bX^{\T}\bxi_u $, yielding a correctly specified OR model of ${\E}_{\mathbb{Q}} \{\bS_{a}(Y, \bZ_{(k)}; \bbeta) \mid \bX \}$.
%The OR model (\ref{eq:HD-OR}) is correctly specified in this setting.
The second uses a nonlinear transformation $\bb(\bX)$ specified in the SM and $m(\bX,\bxi_u) = \bff(\bX)^{\T}\bxi_u$ with $\bff(\bX) = (\bb(\bX)^{\T},X_{11}^2,\ldots,X_{30}^2,X_1,X_5,X_9  )^{\T}$, producing a misspecified OR model.
For both settings, $\bxi_u$ is estimated via Lasso using the $u$th source data.
% $\{1,x_i,x_i^2,x_ix_j : i,j\leq 10; x_{11},\ldots,x_{p} \}$.  
% linear ($X_i$), quadratic ($X_i^2$), and the interaction terms ($X_iX_j$) of  $\{X_1,\ldots,X_{10}\}$, and the remaining part $(X_{11},\ldots,X_{p})$.
The high-dimensional calibration problem in (\ref{eq:EL-observe-multi})-(\ref{eq:EL-CB-HD-multi-source}) is implemented by the package {\it CVXR} with optimizer {\it Mosek} in R. 
The augmented outcome regression in (\ref{eq:EL-outcome-multi-source}) is solved with $\ell_1$-penalized estimation, where the regularization parameter is selected by 10-fold cross-validation. 

    \begin{figure}[htbp]
    \centering
    \includegraphics[width=\linewidth]{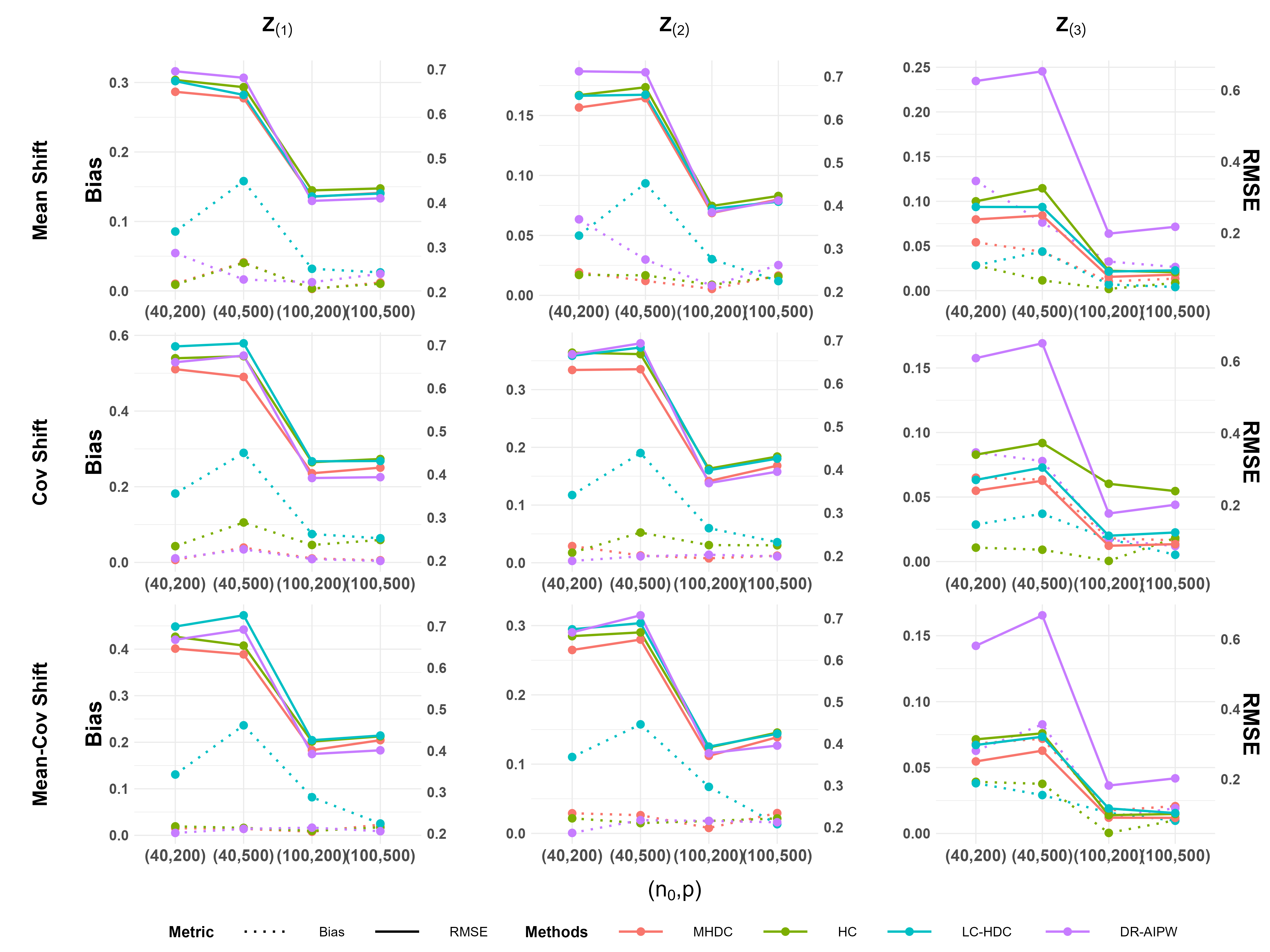}
     \caption{ Absolute Bias (left $y$-axis) and RMSE (right $y$-axis) of the estimated coefficient of $X_3$ using correctly specified OR model for $\E(\bS_a(Y,\bZ;\bbeta)\vert \bX)$ under different settings of covariate shift, $\bZ_{(k)}$, target sample size $n_0 = 40,100$, and dimension $p = 200, 500$.}
    \label{fig: true OR bias boxplot}
    \end{figure}

    Figure \ref{fig: true OR bias boxplot} reports the absolute averaged bias and RMSE of the estimated coefficient of $X_3$ by the proposed MHDC and three competing methods using correctly specified OR model for $\E(\bS_a(Y,\bZ;\bbeta)\vert \bX)$ over 500 repetitions under different covariate shift patterns and subsets of covariates $\bZ_{(k)}$. The left y-axis corresponds to the scale for bias and the right y-axis to that for RMSE.
    Results under the misspecified OR model are shown in Figure S1 in the SM. The MHDC estimator has the smallest RMSEs under almost all settings. Its improvement over LC-HDC validates the efficiency gain in Theorem \ref{thm: efficiency gain of combining all sources}, and its improvement over HC demonstrates the benefit of incorporating the outcome model via PC constraints.
    The MHDC, HC and DR-AIPW estimators show comparably small bias for $\bZ_{(1)}$ and $\bZ_{2}$, whereas LC-HDC exhibits noticeable bias in those settings.
    % The MHDC estimator achieves among the smallest RMSE values, especially under the target covariates $\bZ_{(3)}$, validating the efficiency gain in Theorem \ref{thm: efficiency gain of combining all sources}.
    % Our methods (MHDC, HC, and LC-HDC) remain unbiased regardless of OR model specification, consistent with Theorems \ref{thm: multiply robust consistency of Z estimator}--\ref{thm: IFs of target parameter Multi Source}, while MHDC achieves the minimal RMSE among most cases. 
    % The MHDC estimator achieves the minimal variation compared to HC and LC-HDC, especially under the target covariates $\bZ_{(3)}$, validating the efficiency gain in Theorem \ref{thm: efficiency gain of combining all sources}. 
    % The improvement of MHDC over HC, which incorporates the outcome model via projection calibration constraints, underscores the effectiveness of this approach in enhancing estimation accuracy.
    The DR-AIPW estimator has larger bias and RMSE than MHDC for $\bZ_{(3)}$. Is also shows a significant bias under the misspecified OR model (Figure S1 in the SM), as its DR model is misspecified in the multi-source setting.

Table \ref{table: coverage} reports the average coverage rate of the proposed confidence intervals for $\{\beta_{k,0,j}\}$ in (\ref{eq: confidence interval}) based on the MHDC estimates, where $\beta_{k,0,j}$ is the $j$th component of $\bbeta_{k,0}$.
The coverage rates are close to the nominal level 95\% under a correctly specified OR model.
A slight attenuation occurs under misspecification, but the rates approach the nominal level as the sample size increases, supporting the asymptotic normality of the MHDC estimator.

% From this table, we see that the empirical coverage rates of the proposed confidence intervals are close to the nominal level 95\% when the OR model of $\E(\bS_a(Y,\bZ;\bbeta)\vert \bX)$ is correctly specified. 
% Although a modest attenuation occurs under OR model misspecification, the coverage rates approach the nominal level as the sample size increases, empirically validating the asymptotic normality of the MDEL estimator.

% \usepackage{multirow}
\begin{table}[htbp]
\centering
\resizebox{\textwidth}{!}{
\begin{tabular}{cc|ccccccccc|ccccccccc}
\hline
 &
   &
  \multicolumn{9}{c|}{Correct OR model} &
  \multicolumn{9}{c}{Incorrect OR Model} \\ \hline
\multirow{2}{*}{$n_0$} &
  \multirow{2}{*}{$p$} &
  \multicolumn{3}{c|}{Mean Shift} &
  \multicolumn{3}{c|}{Cov Shift} &
  \multicolumn{3}{c|}{Mean-Cov Shift} &
  \multicolumn{3}{c|}{Mean Shift} &
  \multicolumn{3}{c|}{Cov Shift} &
  \multicolumn{3}{c}{Mean-Cov Shift} \\ \cline{3-20} 
 &
   &
  $\bZ_{(1)}$ &
  $\bZ_{(2)}$ &
  \multicolumn{1}{c|}{$\bZ_{(3)}$} &
  $\bZ_{(1)}$ &
  $\bZ_{(2)}$ &
  \multicolumn{1}{c|}{$\bZ_{(3)}$} &
  $\bZ_{(1)}$ &
  $\bZ_{(2)}$ &
  $\bZ_{(3)}$ &
  $\bZ_{(1)}$ &
  $\bZ_{(2)}$ &
  \multicolumn{1}{c|}{$\bZ_{(3)}$} &
  $\bZ_{(1)}$ &
  $\bZ_{(2)}$ &
  \multicolumn{1}{c|}{$\bZ_{(3)}$} &
  $\bZ_{(1)}$ &
  $\bZ_{(2)}$ &
  $\bZ_{(3)}$ \\ \hline
\multirow{2}{*}{40} &
  200 &
  0.924 &
  0.915 &
  \multicolumn{1}{c|}{0.958} &
  0.929 &
  0.909 &
  \multicolumn{1}{c|}{0.959} &
  0.928 &
  0.915 &
  0.961 &
  0.883 &
  0.887 &
  \multicolumn{1}{c|}{0.932} &
  0.886 &
  0.886 &
  \multicolumn{1}{c|}{0.924} &
  0.887 &
  0.887 &
  0.929 \\
 &
  500 &
  0.904 &
  0.915 &
  \multicolumn{1}{c|}{0.951} &
  0.919 &
  0.919 &
  \multicolumn{1}{c|}{0.956} &
  0.909 &
  0.915 &
  0.951 &
  0.89 &
  0.881 &
  \multicolumn{1}{c|}{0.935} &
  0.899 &
  0.888 &
  \multicolumn{1}{c|}{0.927} &
  0.897 &
  0.887 &
  0.933 \\ \hline
\multirow{2}{*}{100} &
  200 &
  0.915 &
  0.928 &
  \multicolumn{1}{c|}{0.959} &
  0.914 &
  0.927 &
  \multicolumn{1}{c|}{0.961} &
  0.918 &
  0.928 &
  0.961 &
  0.922 &
  0.905 &
  \multicolumn{1}{c|}{0.941} &
  0.924 &
  0.907 &
  \multicolumn{1}{c|}{0.941} &
  0.92 &
  0.911 &
  0.942 \\
 &
  500 &
  0.936 &
  0.926 &
  \multicolumn{1}{c|}{0.962} &
  0.938 &
  0.925 &
  \multicolumn{1}{c|}{0.963} &
  0.939 &
  0.921 &
  0.957 &
  0.897 &
  0.907 &
  \multicolumn{1}{c|}{0.942} &
  0.913 &
  0.907 &
  \multicolumn{1}{c|}{0.935} &
  0.922 &
  0.904 &
  0.94 \\ \hline
\end{tabular}
}
\caption{Average empirical coverage rate of the proposed confidence intervals in (\ref{eq: confidence interval}) using a correct/incorrect OR model for $\E(\bS_a(Y,\bZ;\bbeta)\vert \bX)$ under different settings of covariate shift, $\bZ_{(k)}$, target sample size $n_0 = 40,100$, and dimension $p = 200, 500 $.}
\label{table: coverage}
\end{table}

Table \ref{table: size and power} reports the empirical size and power of the proposed specification test for the transferability hypothesis in (\ref{eq:Hypotheses}). 
The data generation process of the two sources introduced in this section satisfies the null hypothesis of (\ref{eq:Hypotheses}), as the conditional distribution $Y\vert \bX$ is identical for both sources. To evaluate the power, 
we generate $Y = 3.14 + \bX^{\T} {\boldsymbol{ \varpi}}_{a} + \epsilon$ for the second source $\mathbb{F}_2$ with $\boldsymbol{ \varpi}_{a} = (0.5 \cdot \bone_5^{\T}, 1 \cdot \bone_5^{\T}, \bzero_{p-10}) \neq \boldsymbol{ \varpi}$ such that the conditional distribution of $Y$ given $\bX$ differs between the two sources.
% under $\mathbb{F}_2$ is different from that under $\mathbb{F}_1$. 
% We observe that the empirical size stays close to the desired significance level, and the power rises with more covariates in $\bZ$ and larger sample size.
% Our empirical results demonstrate that the test maintains size control, with empirical rejection rates closely approximating the nominal significance level ($\alpha = 0.05$).
% The statistical power exhibits monotonic improvement with increasing dimension of the covariates $\bZ$ or the growing sample size $n_0$.
The empirical sizes are close to the nominal level 0.05 for all cases, indicating good size control. The power increases with sample size $n_0$ and the dimension of target covariates $\bZ_{(k)}$, approaching 1 when $n_0 = 100$.

\begin{table}[htbp]
\resizebox{\textwidth}{!}{
\begin{tabular}{cc|ccccccccc|ccccccccc}
\hline
 &
   &
  \multicolumn{9}{c|}{Size} &
  \multicolumn{9}{c}{Power} \\ \hline
\multirow{2}{*}{$n_0$} &
  \multirow{2}{*}{p} &
  \multicolumn{3}{c|}{Mean Shift} &
  \multicolumn{3}{c|}{Cov Shift} &
  \multicolumn{3}{c|}{Mean-Cov Shift} &
  \multicolumn{3}{c|}{Mean Shift} &
  \multicolumn{3}{c|}{Cov Shift} &
  \multicolumn{3}{c}{Mean-Cov Shift} \\ \cline{3-20} 
 &
   &
  $\bZ_{(1)}$ &
  $\bZ_{(2)}$ &
  \multicolumn{1}{c|}{$\bZ_{(3)}$} &
  $\bZ_{(1)}$ &
  $\bZ_{(2)}$ &
  \multicolumn{1}{c|}{$\bZ_{(3)}$} &
  $\bZ_{(1)}$ &
  $\bZ_{(2)}$ &
  $\bZ_{(3)}$ &
  $\bZ_{(1)}$ &
  $\bZ_{(2)}$ &
  \multicolumn{1}{c|}{$\bZ_{(3)}$} &
  $\bZ_{(1)}$ &
  $\bZ_{(2)}$ &
  \multicolumn{1}{c|}{$\bZ_{(3)}$} &
  $\bZ_{(1)}$ &
  $\bZ_{(2)}$ &
  $\bZ_{(3)}$ \\ \hline
\multirow{2}{*}{40} &
  200 &
  0.04 &
  0.07 &
  \multicolumn{1}{c|}{0.062} &
  0.06 &
  0.076 &
  \multicolumn{1}{c|}{0.08} &
  0.052 &
  0.05 &
  0.034 &
  0.502 &
  0.83 &
  \multicolumn{1}{c|}{0.998} &
  0.426 &
  0.744 &
  \multicolumn{1}{c|}{0.99} &
  0.614 &
  0.864 &
  1 \\
 &
  500 &
  0.042 &
  0.066 &
  \multicolumn{1}{c|}{0.058} &
  0.06 &
  0.066 &
  \multicolumn{1}{c|}{0.056} &
  0.054 &
  0.066 &
  0.024 &
  0.502 &
  0.836 &
  \multicolumn{1}{c|}{0.996} &
  0.352 &
  0.718 &
  \multicolumn{1}{c|}{0.984} &
  0.556 &
  0.854 &
  0.994 \\ \hline
\multirow{2}{*}{100} &
  200 &
  0.028 &
  0.038 &
  \multicolumn{1}{c|}{0.052} &
  0.048 &
  0.052 &
  \multicolumn{1}{c|}{0.052} &
  0.052 &
  0.044 &
  0.062 &
  0.896 &
  0.994 &
  \multicolumn{1}{c|}{1} &
  0.876 &
  1 &
  \multicolumn{1}{c|}{1} &
  0.896 &
  0.996 &
  1 \\
 &
  500 &
  0.038 &
  0.042 &
  \multicolumn{1}{c|}{0.03} &
  0.064 &
  0.064 &
  \multicolumn{1}{c|}{0.086} &
  0.064 &
  0.086 &
  0.086 &
  0.894 &
  1 &
  \multicolumn{1}{c|}{1} &
  0.872 &
  0.994 &
  \multicolumn{1}{c|}{1} &
  0.888 &
  0.992 &
  1 \\ \hline
\end{tabular}
}
\caption{
% Empirical Power of the Specification Test under different scenarios of covariate shift with $n_0 = 40,100$ and $p = 200,500 $ 
Empirical size and power of the specification test for the hypothesis in (\ref{eq:Hypotheses}) under different settings of covariate shift, $\bZ_{(k)}$, $n_0$, and $p$. The nominal level is $\alpha = 0.05$.
}
\label{table: size and power}
\end{table}

	\setcounter{equation}{0}
	\section{Real Data Analysis}\label{sec: real data}

As introduced in Section 2, we aimed to study the relationship between ozone concentration and two atmospheric drivers: total solar radiation (TSR) and temperature (TEMP). 
% As China industrialized with much increased anthropogenic activities, north China plain region has experienced severe air pollution in the last two decades. While the primary air pollutant, the airborne particular matters (PM), are coming down in North China due to the implementation of environmental protection and air quality-related policies, ground level $\text{O}_3$ is on the rise \citep{Li2019Anthropogenic}. A recent study has established the relationship between the  $\text{O}_3$ concentration and a variety of meteorological variables \citep{Li2021radiative}.
% such as the surface total solar radiation, air temperature and relative humidity \citep{Li2021radiative}.
% In this section, the proposed method is applied to investigate the relationship between  $\text{O}_3$ concentration and meteorological variables.
% We focus on four major cities in North China: Beijing, Tianjin, Jinan and Taiyuan, where Beijing is treated as target domain. 
The scope of our study was the spring (March 1st to May 31st) in 2013, as the ozone level in spring is generally high. 
The hourly monitored $\text{O}_3$ concentration from the China Meteorological Administration (CMA) monitoring network was considered as the response variable.
The covariates $\bX$ included 
% hourly $\text{PM}_{2.5}$, $\text{PM}_{10}$, nitrogen dioxide ($\text{NO}_2$), 
cumulative wind speed in four directions (INE, INW, ISE, ISW), surface total solar radiation (TSR), surface air temperature (TEMP), relative humidity (HUMI), boundary layer height (BLH), the low (LCC), medium (MCC) and high (HCC) cloud cover percentages, as well as their one, two, three hours-lagged terms and quadratic terms. 
The cumulative wind speed data were obtained from China Environmental Monitor Center sites in each city, the TSR data were collected from CMA, and the others were obtained from the European Center for Medium-Range Weather Forecasts. 
% A DAY variable counting the number of days since March 1st was also included for its statistical significance in modeling the $O_3$ proposed in \citep{Li2021radiative}.

We were interested in the regression coefficients of $\text{TSR}_{t-3}$, $\text{TEMP}_{t-3}^2$, and $\text{TSR}_{t-3}^2$ for the $\text{O}_3$ concentration in Xi'an, where the subscripts denote the time-lag. \cite{Li2021radiative} reported that those three covariates are the top three significant variables related to $\text{O}_3$.
The absence of $\text{O}_3$ measurements in Xi'an during the spring of 2013 necessitates the use of source datasets to conduct the analysis.
% The data collection started in 2013, resulting in low-quality and systematically missing observations during that spring. 
We adopted the data from Tianjin, Jinan, and Taiyuan as multiple sources, where the $\text{O}_3$ measurements were available during the study time. Those three cities had a comparable level of air pollution to the target city, Xi'an.
We applied the proposed methods MHDC and LC-HDC, the existing method DR-AIPW 
considered in the simulation study, and the estimated regression coefficients on the combined source data ($\mbox{Est}_{\text{s}}$) without considering covariate shift. 

\begin{table}[htbp]
\centering
\resizebox{0.6\textwidth}{!}{
\begin{tabular}{c|c|c|c|c|c}
& MHDC   & LC-HDC & DR-AIPW   & $\mbox{Est}_{\text{s}}$   \\ \hline
$\text{TSR}_{t-3}$    & 1.091(0.109) & 0.749(0.114)  & 0.898    & 0.912 \\ \hline
$\text{TEMP}_{t-3}^2$ & 0.308(0.042)  & 0.305(0.048)  & 0.396    & 0.367 \\ \hline
$\text{TSR}_{t-3}^2$  & -0.535(0.119) & -0.252(0.122)& -0.432  & -0.43
\end{tabular}
}
\caption{Estimated coefficients (estimated standard error for MHDC and LC-HDC estimators) of $\text{TSR}_{t-3}$, $\text{TEMP}_{t-3}^2$ and $\text{TSR}_{t-3}^2$ on $\text{O}_3$ concentration in the spring of 2013 at Xi'an }
\label{table: xi an 2013}
\end{table}

The result in table \ref{table: xi an 2013} reveals the effect of each target variable. TSR with a 3-hour lag and its quadratic term exhibited a nonlinear relationship with ozone concentration: a moderate TSR increase elevated $\text{O}_3$ levels, whereas a further increase of TSR reduced the rate of ozone formation. The estimated coefficients of the temperature were similar across the four methods, demonstrating a nonlinear promoting effect.
% except for the LC-DEL estimator, but all demonstrated a nonlinear promoting effect.
This phenomenon was consistent with the photochemical mechanisms \citep{WangW2019AerosolsandOzone} and the results established in \citet{Li2021radiative}.
% wherein certain TSR and TEMP levels contribute to increased ozone concentration \citep{WangW2019AerosolsandOzone}.
% add std in table

% time of evaluation

\setcounter{equation}{0}
\section{Discussion}\label{sec: conclusion}

In this paper, we propose a new method for statistical inference of parameters in a target population with missing response variables by carefully designing calibration constraints to utilize data from multiple sources and achieve Neyman-orthogonality under high-dimensional covariates. 
%novel targeted learning estimator, MDEL (Multi-source Debiased Empirical Likelihood), for estimating parameters of interest in the target population with missing observations under high-dimensional covariate shift scenarios and multiple source domains.  
The proposed method combines multiple DR and OR models, ensuring valid statistical inference against model misspecification. Moreover, it achieves the most efficient estimation in the multiple-source settings.

Our method connects to several existing works in missing data and causal inference but brings several new perspectives. Compared to covariate balancing methods \citep{Imai2014CBPS, Zubizarreta2015} in fixed or low dimensional settings, our approach extends those methods to high-dimensional settings via adding a set of augmented constraints, which include outcome model projection constraint and orthogonality constraints to construct Neyman orthogonal score functions for the target parameters.
%that balance the derivatives of working DR and pre-trained response regression models.
Compared to HD AIPW methods \citep{Tan2020Modelassited, Ning2020Robust}, we develop a unified framework that combines multiple sources and multiple OR/DR models, 
%achieving first-order insensitivity to nuisance estimation 
while retaining a simple IPW formulation. 
%and avoiding the potential non-convexity of AIPW estimators. 
%Compared with recent data-fusion studies like \cite{Li2023Datafusion}, which derive general efficiency bounds under an abstract data-generating process without target data, our framework targets practical problems with observed targets and missing outcomes, emphasizing how information from multiple sources and models can be jointly leveraged. 
In a summary, the proposed method brings the following new perspectives to the existing literature: (1) it uses a simple IPW formulation to achieve multiply robust inference under high-dimensional covariates; (2) it facilitates an easy solution for M-estimation and estimating equations without incurring the potential nonconvex issue caused by the AIPW formulation; (3) it provides a new framework for establishing Neyman orthogonality for missing data and causal inference problems; and (4) it demonstrates estimation efficiency for data from multiple sources.

%Through comprehensive theoretical analysis and empirical validation, we demonstrate that MDEL effectively addresses the challenges posed by multi-source high-dimensional covariate shifts while maintaining statistical efficiency and robustness.

The proposed method is based on the assumption of transferability, namely, the covariate shift assumption defined in Condition \ref{as:ignorability}. For multi-source data, the transferability assumption consists of two parts. One is the equality of the conditional distribution of $Y$ given $\bX$ across different source samples. The other is the equality of the conditional distribution between the source and target samples. The first part of the transferability assumption is testable based on observed data. However, directly testing a high-dimensional conditional distribution is difficult. This motivates us to consider the hypotheses in (\ref{eq:Hypotheses}) for the equality of the limiting values of the estimated parameters using different sources of data in Section \ref{sec:test}. The second part of the transferability assumption cannot be verified from the samples at hand. If the second part of the transferability assumption is questionable, sensitivity analysis can be considered \citep{Cinelli2020Sensitivity,LuandDing2024}.
%\textcolor{red}{(I wonder whether we can distinguish between the transferability from the source sample to target sample and the homogeneity among the source samples.) }

The setting of multi-source is naturally related to federated learning, where the data in each source site are computed locally and only summary statistics are passed to the target site due to computation or data privacy constraints. Note that the proposed single-source HDC method can be conducted on each source site separately, using summary statistics from the target site. This means the linear combination estimator in (\ref{eq:combined-est}) can be applied to the federated learning setting. However, the MHDC estimation in (\ref{eq:obj-HD-multi}) requires pooling all source data together, which is not directly applicable to the federated learning setting.

%\section*{Supplementary Materials}

%The supplementary material provides additional simulation results, definitions of the probability limiting values of the nuisance parameters, and the proofs for Theorem \ref{thm: multiply robust consistency of Z estimator} to \ref{thm: efficiency gain of combining all sources} and Proposition \ref{prop: specification test}.

%\clearpage
\bibliographystyle{apalike}
\bibliography{reference}

\end{document}